\documentclass[prb,twocolumn,amsmath,amssymb,superscriptaddress,citeautoscript,longbibliography]{revtex4-2}

\usepackage{graphicx,xcolor}
\usepackage{lipsum}
\usepackage{braket}
\usepackage[colorlinks,urlcolor=blue,citecolor=blue]{hyperref}
\usepackage[utf8]{inputenc}
\usepackage{verbatim}
\allowdisplaybreaks

\begin{document}

\title{Systematic compactification of the two-channel Kondo model. \\III.
\textit{Extended field-theoretic renormalization group analysis}}

\author{Aleksandar \surname{Ljepoja}}
\affiliation{Department of Physics, University of Cincinnati, OH-45221, USA}
\author{C.~J.~\surname{Bolech}}
\affiliation{Department of Physics, University of Cincinnati, OH-45221, USA}
\author{Nayana \surname{Shah}}
\affiliation{Department of Physics, Washington University in St.~Louis, MO-63160, USA}

\begin{abstract}
We carry out a field-theoretical renormalization group procedure based on the Callan-Symanzik equation to calculate the detailed flow for the (multi) two-channel Kondo model and its compactified versions. In doing so, we go beyond the universal terms in the beta function we obtained using poor man's scaling (see \href{https://doi.org/10.48550/arXiv.2308.03590}{companion paper II}) and culminate our analysis of how the compactified versions of the model fare against the original one. Among other results, we explore the large-channel-number limit and extend our considerations to the finite temperature crossover region. Moreover, we gain insights into the contradistinction between the consistent vs.~conventional bosonization-debosonization formalisms, thereby advancing our understanding on multiple fronts. In particular, we make use of renormalization-flow arguments to further justify the consistent re\-fermionization of the parallel Kondo interaction we presented earlier (see \href{https://doi.org/10.48550/arXiv.2308.03569}{companion paper I}). 
\end{abstract}

\maketitle

\section{Introduction}

The initial development of the renormalization group (RG) is intimately related to the efforts to tame the infinities that appear in many of the perturbative calculations based on quantum (field) theory; ---one of the earliest examples of which goes back to the determination of the Lamb shift between the $2s$ and $2p$ levels of the hydrogen atom \cite{Bethe1947}. In the early 1950's, St\"uckelberg and Petermann \cite{Stueckelberg1953} introduced the notion of infinitesimal Lie group transformations generated by operators connected to the renormalization of the electron charge. Subsequently, parallel work by Gell-Mann and Low \cite{GellMann1954} and by Bogoliubov and Shirkov \cite{Bogoliubov1956} derived functional equations for the quantum-electrodynamics propagators and vertices in the general massive case. 

About a decade later, the subsequent generalization to arbitrary $n$-point correlation functions for general quantum field theories and the connection with the ideas of scaling led to the RG (differential) equations associated with the work of Callan \cite{Callan1970} and Symanzik \cite{Symanzik1970,*Symanzik1971}. Around the same time, the field-theoretic RG methods were being applied to the Kondo Hamiltonian \cite{Abrikosov1970,Fowler1971} to understand its flow to strong coupling at low temperatures. This flow was also being studied based on mappings to the 1D Coulomb-gas problem and the inverse-square Ising model, and via a direct scaling of the $T$-matrix (in a series of works by Anderson and collaborators culminating with the poor man's scaling (PMS) proposal \cite{Anderson1969,*Yuval1970,*AndersonYH1970,*Anderson1970} that preluded the development of the Wegner-Wilson momentum-shell RG \cite{Wegner1972,*Wegner1973,Wilson1974}, the convergence with the scaling ideas from the study of critical phenomena, and the development of the numerical renormalization group (NRG) for the Kondo problem \cite{Wilson1975}). The PMS approach is directly equivalent to the field-theoretic one when taken to the first order in the coupling expansion, but a careful wave-function renormalization of the initial and final $T$-matrix states is required to extend the equivalence to the next order \cite{Solyom1974}. To even higher orders, the PMS procedure loses its alluring simplicity, and the field-theoretic method based on the Callan-Symanzik (CS) equation provides a more systematic and general approach. Moreover, the CS equation makes it clear how to perform higher-order logarithmic-divergence re\-summations (the so-called \textit{leading-logs}, \textit{next-to-leading-logs}, and so on) \cite{Shirkov2001}.

In the previous work of this series, we used a bosonization-de\-bosonization (BdB) procedure proposed earlier \cite{shah2016,*bolech2016} to show how to systematically derive compactified versions of Kondo-type impurity models, and we performed some initial comparisons in certain exact limits of the original (multi) two-channel Kondo model and two different compatified versions of it \cite{Ljepoja2024a}. We further showed as well that their RG flows are exactly equivalent to the level captured by PMS \cite{Ljepoja2024b}. We have indications, however, that such equivalence is lost at higher orders (as has to be the case, since exact comparisons in special-case limits yield differing results, especially for the nonequilibrium transport characteristics \cite{Ljepoja2024a}). Exploring those differences is the purpose of the present article, and it requires a more systematic RG study than what standard PMS affords. It can thus be better undertaken using a field-theoretic approach based on the CS equations. That method has been notably used in the context of Kondo-type models to study their ``large-$M$'' perturbative solution \cite{gan1993,gan1994,Shaw1998,bensimon2006} and to address nonequilibrium steady-state regimes \cite{doyon2006,Culver2021}, both of which will be seen to be physically interesting comparisons beyond the general-case considerations.

The rest of our work is organized as follows. In the next section we present some general aspects of the field-theoretic RG method starting from specific auxiliary considerations pertaining to spin systems (needed to handle the Kondo impurity) and followed by a presentation of the CS equation. The two subsequent sections detail the RG flow calculations first for the original model or \textit {direct scheme} (recovering the PMS results), and then for the compactified models referred as \textit{indirect schemes}. In Sec.~\ref{Sec:Beyond} the results are extended beyond the third order in the beta function; differences arise and a detailed comparison is given. In Sec.~\ref{Sec:extRG} we present an extended version of the RG analysis that further clarifies the differences between ``schemes''. In the last two sections we present alternative arguments for the re\-fermionization of the parallel Kondo interaction (used during the compactification of the model) and we conclude by discussing and providing highlights of the main findings. Several appendixes provide further details and additional results.

\section{Field-Theoretical RG Procedure}

The multichannel Kondo model with spin and channel anisotropy is given by a Hamiltonian of the form
\begin{equation}
\begin{split}\label{eq:direct}
    H &= H_{0}+H^{\perp}_{K}+H^{z}_{K}\\
    H_{0} &= \sum_{\sigma \ell \alpha}\int dx \ \psi^{\dagger}_{\sigma \ell \alpha}(x,t) (iv_{F}\partial_{x})\psi^{}_{\sigma \ell \alpha}(x,t) \\
    H^{z}_{K} &= \sum_{\sigma \ell \alpha} \sigma J_{\ell \ell}^{z} S^{z} \psi^{\dagger}_{\sigma \ell \alpha}(0,t)\psi_{\sigma\ell \alpha}(0,t)\\
    H^{\perp}_{K} &= \sum_{\sigma \ell \alpha} J_{\ell \ell}^{\perp} S^{\sigma} \psi^{\dagger}_{\bar{\sigma} \ell \alpha}(0,t)\psi_{\sigma\ell \alpha}(0,t)
\end{split}
\end{equation}
where each electron carries two nonspin ``flavor'' labels (that combined give the total number of channels $K=2M$), $\ell = \{L,R\}=\{-, +\}$, and $\alpha = 1 , \ldots , M$. In principle, this model can be thought of as the two-channel model where each of the two channels has an additional $M$-fold degeneracy. We also have a spin index, $\sigma=\{\downarrow, \uparrow\} =\{-, +\}$, with the notation $\bar{\sigma}=-\sigma$. In general, the exchange coupling constant can be both spin- and channel-anisotropic. Channel anisotropy in a restricted sense will mean that the coupling constants of each $\ell$ channel with the impurity spin are different while the $\alpha$-degeneracy is not broken. However, we are going to be working (mostly) with the spin-anisotropic and channel-isotropic model.

\subsection{Popov-Fedotov Parameterization}
In order to be able to carry out efficient perturbative (diagrammatic) calculations for the model, it is better to rewrite the impurity spin operators in terms of fermionic degrees of freedom. In that way Wick's theorem is applicable \cite{Wick1950}; otherwise, one is unable to use the standard diagram technique. A way this can be done is by employing the Popov-Fedotov (PF) method \cite{popov1988}, in which we rewrite the spin degrees of freedom in terms of pseudofermions
\begin{equation}
    S_\mathrm{imp}=\sum_{\mu, \nu} \eta^{\dagger}_{\mu} \sigma_{\mu \nu} \eta_{\nu}
\end{equation}
Here the $\eta$'s are fermionic degrees of freedom, but with an imaginary chemical potential so that the empty and doubly occupied impurity spin states are automatically excluded. We refer to them as pseudofermions partly because they do not have a kinetic-energy contribution to the Hamiltonian, --as well as to distinguish them from the actual (electron) fermionic degrees of freedom in the model--, but for all intent and purposes in the construction of Feynman diagrams they are treated just like any other fermionic degree of freedom. This spin-handling technique is described in further detail in Appendix \ref{Apdx:PF}. 

Having the spin operators rewritten in terms of the PF pseudo\-fermions enables us to use a standard fermionic-path-integral formulation of the model, in which the free part of the action, written in terms of Grassmann variables, is now given by
\begin{equation}
\begin{split}
S_{0} =  & \ \rho_{0} \sum_{\sigma \ell \alpha}\int_{-D}^{D} d\epsilon \int_{0}^{\beta} d \tau \bar{\psi}_{\sigma \ell \alpha}(\epsilon, \tau)(\epsilon +\partial_{\tau}) \psi_{\sigma \ell \alpha}(\epsilon, \tau) \\
& \qquad \qquad + \sum_{\sigma} \int_{0}^{\beta} d \tau \bar{\eta}_{\sigma}(\tau) (\mu + \partial_{\tau})  \eta_{\sigma}(\tau)\\
\end{split}
\end{equation}
where $\rho_{0}$ is the one-dimensional density of states (per spin, per channel), $D$ is the energy cutoff for the fermions (the bandwidth) in the model, and we introduced the pseudofermion imaginary chemical potential, $\mu = - i \pi / 2\beta$, as per the PF prescription.
We also performed a Fourier transformation from position- to momentum-space representations, and replaced the summations over momenta by integrals in energy:
\begin{equation}
\frac{1}{V}\sum_{k} \square \quad \mapsto \quad 
\rho_{0} \int_{-D}^{D} \square \, d\epsilon
\end{equation}

On the other hand, we have kept the action formulated in imaginary time, $\tau$, rather than switching to (Matsubara) frequencies as usual. The reason for such a choice is the relative simplicity of carrying out perturbative diagrammatic calculations in this way. Indeed, integrals rather than frequency summations present fewer subtleties when multiple nested ones need to be evaluated at high orders in perturbation theory. 

The interacting part of the action is described by the Kondo-exchange terms
\begin{widetext}
\begin{equation}
S_{I} = \rho_{0}^2 \int_{-D}^{D} d\epsilon \, d\epsilon^{\prime}  \int_{0}^{\beta} d \tau \   
\left[ J_{\perp}\sum_{\sigma \ell \alpha} \bar{\eta}_{\sigma}(\tau) \eta_{\bar{\sigma}}(\tau) \bar{\psi}_{\bar{\sigma} \ell \alpha}(\epsilon, \tau) \psi_{\sigma \ell \alpha}(\epsilon^{\prime} , \tau)
+ J_{z}\sum_{\sigma \sigma^{\prime}}\sum_{\ell \alpha} \sigma \sigma^{\prime} \bar{\eta}_{\sigma^{\prime}}(\tau)  \eta_{\sigma^{\prime}}(\tau) \bar{\psi}_{\sigma \ell \alpha}(\epsilon, \tau) \psi_{\sigma \ell \alpha}(\epsilon^{\prime}, \tau) \right]
\end{equation}
\end{widetext}

Given $S_{0}$, the noninteracting Green's functions for fermions and pseudofermions are given by
\begin{equation} \label{eq:propagators}
\begin{split}
G^\mathrm{f}_{0}(\tau,\epsilon) & =e^{-\tau \epsilon} \bigg[\theta(\tau) \big (1-n_{F}(\epsilon) \big ) +\theta(-\tau) n_{F}(\epsilon)\bigg]\\
G^\mathrm{pf}_{0}(\tau,\mu)&=e^{-\tau \mu} \bigg[\theta(\tau) \big (1-n_{F}(\mu) \big ) +\theta(-\tau) n_{F}(\mu)\bigg]\\
\end{split}
\end{equation}
and one can use perturbative calculations to incorporate the effects of $S_{I}$ in any physical quantity of interest.

\subsection{Callan-Symanzik Equation}
For the purpose of finding the RG flow of the Kondo couplings, we choose to focus on the electron self-energy. One reason for this is that the self-energy is the simplest vertex function of the theory, and there is a convenient series expansion that allows its immediate computation. (Moreover, we already performed exact comparisons of the self-energy in the flat-band limit for the two compactifications being compared here  \cite{Ljepoja2024a} and know that they are different.) We shall actually be interested in the imaginary part of the retarded self-energy, which is directly related to the scattering rate of the fermions by the impurity and is obtained via an analytic continuation of the perturbatively calculated Matsubara self-energy. As such, it is a physical observable and thus expected to be scale invariant. This allows us to obtain the RG flow by using the CS equation, which in its generic form is given by the total derivative of the observable \cite{Schwartz}
\begin{equation}
    \frac{d \Sigma^{i}}{d \mathrm{\ln}(D)} = \gamma_{s} \Sigma^{i}
    \label{eq:CS}
\end{equation}
where $\gamma_{s}$ is the corresponding \textit{anomalous scaling dimension} of $\Sigma^{i}$. Importantly, the latter should be identically zero for the case of a direct physical observable (independent of field normalization and typically associated to a macroscopic symmetry) \cite{Goldenfeld1989,PeskinSchroeder}. Here $D$ is the cutoff of the model already introduced above but entering the equation explicitly via its differential logarithm (thus, it could be divided by another arbitrary energy scale in order to make it dimensionless, a natural choice for that reference scale would be the frequency at which $\Sigma^{i}$ is evaluated). As we shall encounter below, the CS equation will translate into a linear system of equations involving the known coefficients of the expansion of the self-energy in powers of the coupling constant and the unknown coefficients in the respective expansions of the beta functions. As it turns out, this typically provides an over\-determined but consistent system of equations (in all schemes not just the \textit{direct} one) that can be solved to find the RG flow of the model encoded in the beta functions. Such calculations are presented in the following several sections.

\section{RG calculation in the Direct Scheme}

In this section we illustrate the methodology for obtaining the RG flow (beta functions) by using the CS equation. We shall do it here for the original model (which we refer to as the \textit{direct scheme}) and leave the discussion of the calculation based on the compactified versions of the model (or \textit{indirect schemes}) for subsequent sections.

\subsection{Electron Self-energy}

The first step is to obtain the imaginary part of the electron self-energy which we choose as our scale-invariant observable. Although one does not have access to the exact self-energy, it is possible to calculate it iteratively by using the perturbative diagrammatic technique. The expansion is done in the powers of the Kondo-exchange coupling constants. In the diagrams to be presented below, a full line indicates the electron propagator and a dashed line corresponds to a pseudo\-fermion one; the expressions for both were given in Eq.~(\ref{eq:propagators}). Each vertex carries one power of the coupling constant: $J_{\perp}$ for the spin-flip vertex and $J_{z}/2$ for the no-spin-flip one. Diagrams are shifted in imaginary time in such a way that the leftmost vertex is always at time $\tau_0=0$ and the rightmost at $\tau$, while internal vertices are integrated from $0$ to $\beta$ when evaluating the diagram. Finally, the energy (momentum) of each internal fermion propagator is independent and also has to be integrated, in this case over the entire bandwidth determined by $D$.

\begin{figure}[t!]
\includegraphics[width=0.47\textwidth]{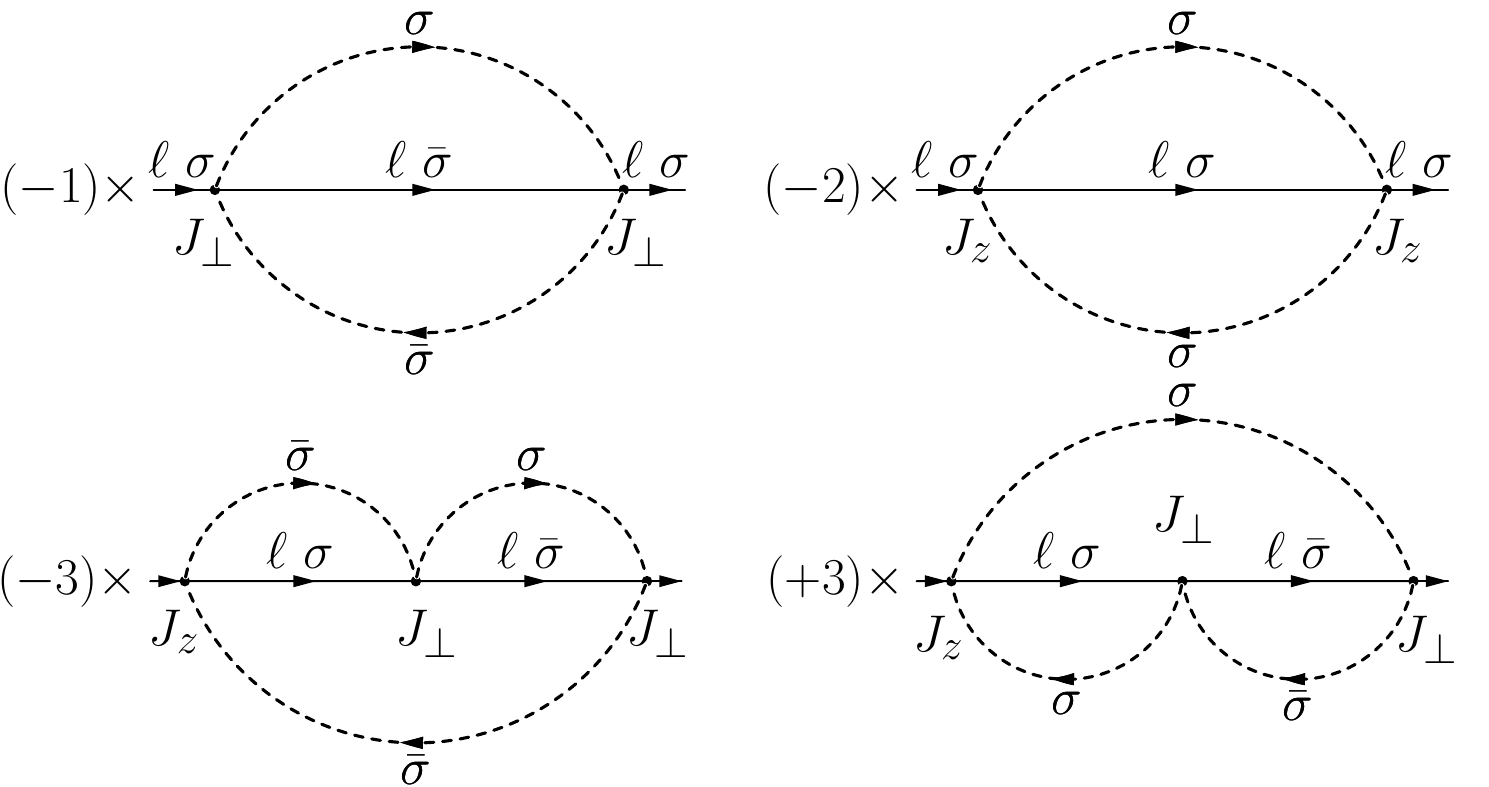}
\caption{Second- and third-order diagrams contributing to the electron self-energy (for a given $\alpha$, not indicated). The Feynman rules are as follows. Dashed lines are pseudo\-fermion propagators, and the solid lines represent regular-fermion ones. The number in front of a diagram is its multiplicative factor. It counts the number of possible vertex permutations, whether the spins are aligned or antialigned on the $J_{z}$ vertex, and the number of possible spin orientations along the closed pseudo\-fermion loop. There is, as well, an extra minus sign coming from the pseudo\-fermion loop itself.}
\label{fig:2nd3rdSE}
\end{figure}

The second- and third-order self-energy diagrams are shown in Fig.~\ref{fig:2nd3rdSE}. Here we explicitly show two versions of the third-order diagram that can be obtained from each other by the reversal of the pseudo\-fermion propagators. We refer to them as particle-hole-related pairs. For some diagrams such pairs do not exist. That is the case of the second-order diagrams, and some of the fourth-order ones that are symmetric with respect to the reversal of pseudo\-fermion line arrows. In subsequent figures, when there is a particle and a hole version of a diagram we will show only the hole one (to keep the figures more succinct). 

We shall now illustrate the calculation of the second- and third-order contributions in order to provide some of the technical details (the calculations proceed in analogous ways for higher-order contributions). 

Let us first consider the $J_{\perp}^{2}$ diagram (top-left in Fig.~\ref{fig:2nd3rdSE}); the Feynman rules give
\begin{equation}
\Sigma_{(2,0)\perp}(\tau)= -J_{\perp}^{2} \rho_{0} \int_{-D}^{D} d\epsilon_{1} G^\mathrm{f}_{0}(\tau,\epsilon_{1})G^\mathrm{pf}_{0}(\tau,\mu)G^\mathrm{pf}_{0}(-\tau,\mu)
\end{equation}
In this diagram there are no internal vertices, and so there are no integrals in imaginary time. There is, however, an internal fermionic propagator with energy $\epsilon_{1}$, which has to be integrated over as indicated. We want to have the self-energy as a function of frequency, $\omega$, because we eventually want to do an analytic continuation in order to extract the imaginary part of the retarded self-energy. Using the Fourier convention $\Sigma_{2,0}(\omega)=\int_{0}^{\beta} d\tau e^{i\omega \tau}\Sigma_{2,0}(\tau)$, and letting $\mu \rightarrow \frac{i \pi}{2 \beta}$, we find that the diagram contribution in frequency space is given by
\begin{equation}
\Sigma_{(2,0)\perp}(\omega)=- J_{\perp}^{2}  \rho_{0} \int_{-D}^{D} d\epsilon_{1}\frac{1}{2(\epsilon_1-i\omega)}
\end{equation}
Before the integration, we are going to carry out the analytical continuation by replacing
\begin{equation}\label{eq:ana_con}
    \frac{1}{(\epsilon_{1} - i \ \omega)} \quad\mapsto\quad \frac{1}{(\epsilon_{1} - \omega)} - i \pi \delta (\epsilon_{1} - \omega)
\end{equation}
We keep only the imaginary part (that involves a delta function and makes integration trivial), giving the second-order $J_{\perp}^2$ contribution to the imaginary part of the self-energy as
\begin{equation}
\Sigma^{i}_{(2,0)\perp}(\omega)=\frac{\pi}{2} J_{\perp}^{2}  \rho_{0}
\end{equation}
This is a constant contribution with no explicitly dependence on $\omega$. The contribution from the $J_{z}^2$ diagram is calculated in the same way, the only difference being an overall multiplicative factor,
\begin{equation}
\Sigma^{i}_{(2,0)z}(\omega)= \frac{\pi}{4} J_{z}^{2}  \rho_{0}
\end{equation}

\begin{widetext}
We switch now our attention to the third-order diagrams (bottom row in Fig.~\ref{fig:2nd3rdSE}). These are somewhat more involved to evaluate than the second-order ones because they have an internal vertex, so one has to first do the corresponding integration in imaginary time. The particle and hole third-order diagrams translate, respectively, to the following expressions:
\begin{equation}
\begin{split}
& \Sigma^{p}_{(3,0)}(\tau) = \frac{3}{2}J^{2}_{\perp} J_{z}  \rho_{0}^2 \int_{-D}^{D} d \epsilon_{1} \int_{-D}^{D} d \epsilon_{2} \int_{0}^{\beta} d \tau_{1} \ G^{f}_{0}(\tau_{1},\epsilon_{1})G^{f}_{0}(\tau-\tau_{1},\epsilon_{2})G^{pf}_{0}(\tau,\mu)G^{pf}_{0}(\tau_{1}-\tau,\mu)G^{pf}_{0}(-\tau_{1},\mu)\\
& \Sigma^{h}_{(3,0)}(\tau) =- \frac{3}{2} J^2_{\perp} J_{z}  \rho_{0}^2 \int_{-D}^{D} d \epsilon_{1} \int_{-D}^{D} d \epsilon_{2} \int_{0}^{\beta} d \tau_{1} \ G^{f}_{0}(\tau_{1},\epsilon_{1})G^{f}_{0}(\tau-\tau_{1},\epsilon_{2})G^{pf}_{0}(\tau_{1},\mu)G^{pf}_{0}(\tau-\tau_{1},\mu)G^{pf}_{0}(-\tau,\mu)
\end{split}
\end{equation}
We proceed by doing the integral in $\tau_{1}$ to obtain expressions for $\Sigma^{p}_{3,0}(\tau)$ and $\Sigma^{p}_{3,0}(\tau)$ that we can then Fourier transform in $\tau$. 
(If the perturbation is extended to higher orders, the number of integrals grows accordingly.) 
The resulting particle and hole frequency-dependent contributions are
\begin{equation}
\begin{split}
& \Sigma^{p}_{(3,0)}(\omega) =\frac{3}{2} J^{2}_{\perp} J_{z}  \rho_{0}^2 \int_{-D}^{D} d\epsilon_{1} \int_{-D}^{D} d\epsilon_{2} \bigg [\frac{-i+\tanh(\frac{\beta \epsilon_{2}}{2})}{2(\epsilon_{1}-\epsilon_{2})(\epsilon_{1}-i\omega)} \bigg ] \\
& \Sigma^{h}_{(3,0)}(\omega) = -\frac{3}{2} J^{2}_{\perp} J_{z}  \int_{-D}^{D} d\epsilon_{1} \int_{-D}^{D} d\epsilon_{2} \bigg [\frac{(\frac{1}{2}+\frac{i}{2})(i+e^{\beta \epsilon_{2}})}{(1+e^{\beta \epsilon_{2}})(\epsilon_{1}-\epsilon_{2})(\epsilon_{1}-i\omega)} \bigg ]
\end{split}
\end{equation}
where we also replaced $\mu \rightarrow \frac{i \pi}{2 \beta}$, as per the PF prescription for spin-1/2.
The particle and hole contributions are now added before the analytic continuation is done. We arrive at a relatively simple expression,
\begin{equation}
\begin{split}
\Sigma_{(3,0)}(\omega)& =\Sigma^{p}_{(3,0)}(\omega) + \Sigma^{h}_{(3,0)}(\omega)
=  \frac{3}{2} J_{\perp}^2 J_{z}  \rho_{0}^2 
\int_{-D}^{D} d \epsilon_{1} \int_{-D}^{D} d \epsilon_{2} 
\bigg[ \frac{\tanh(\frac{\beta \epsilon_{2}}{2})}
{(\epsilon_{1}-\epsilon_{2})(\epsilon_{1}-i\omega)} \bigg]
\end{split}
\end{equation}
\end{widetext}
and can proceed with the analytic continuation, using Eq.~(\ref{eq:ana_con}), and the extraction of the imaginary part. We finally do the energy integrals to obtain
\begin{equation}\label{eq:3rdSE}
\begin{split}
\Sigma^{i}_{(3,0)}(\omega)=\pi \frac{3}{2} J_{\perp}^2 J_{z}  \rho_{0}^2 \bigg[ \ln \bigg( \frac{D}{\omega}-1 \bigg) + \ln \bigg( \frac{D}{\omega}+1 \bigg) \bigg]
\end{split}
\end{equation}
Of note is that the integration in energy was done in the zero-temperature limit, in which the numerator of the integrand becomes a sign function. Since we will assume that $D$ is large and $\omega$ small, we can expand the term in the brackets in powers of $\tilde{\omega} \equiv \omega /D \ll 1$, and we  keep only the leading (divergent) logarithmic contribution
\begin{equation}
\begin{split}
\Sigma^{i}_{(3,0)}(\omega)& =-3 \pi J_{\perp}^2 J_{z}   \rho_{0}^2 \mathrm{\ln} (\tilde{\omega}) 
\end{split}
\end{equation}
This is the third-order contribution to the imaginary part of the electron self-energy. It is the first order at which logarithmic divergencies start to appear. In calculating this, and in all other higher-order terms, we will be especially interested in the constant and logarithmic\-ally divergent terms. (In a later section, we shall discuss in detail what happens if additional terms are also taken into account.)

\begin{figure}
\includegraphics[width=0.3\textwidth]{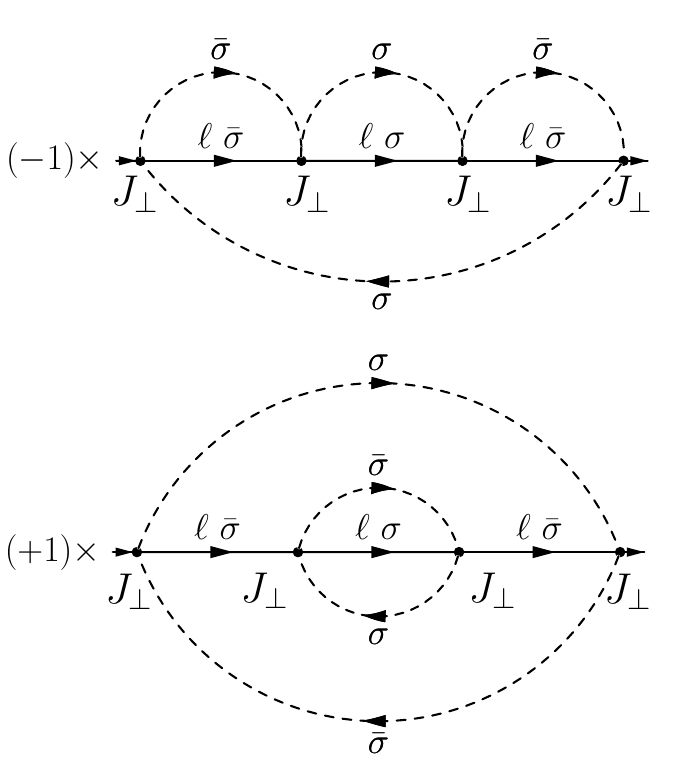}
\caption{Fourth-order diagrams contributing to the $J^{4}_{\perp}$ term in the self-energy. We only show hole diagrams in the figures to keep them more compact. (Of course, some diagrams such as the bottom one shown here with two pseudofermion loops nested in each other are symmetric with respect to the reversal of arrows and therefore have no particle and hole distinction.) We give multiplicative factors in front of each diagram.}
\label{fig:4thXloopSE}
\end{figure}

\begin{figure*}[t]
\includegraphics[width=0.95\textwidth]{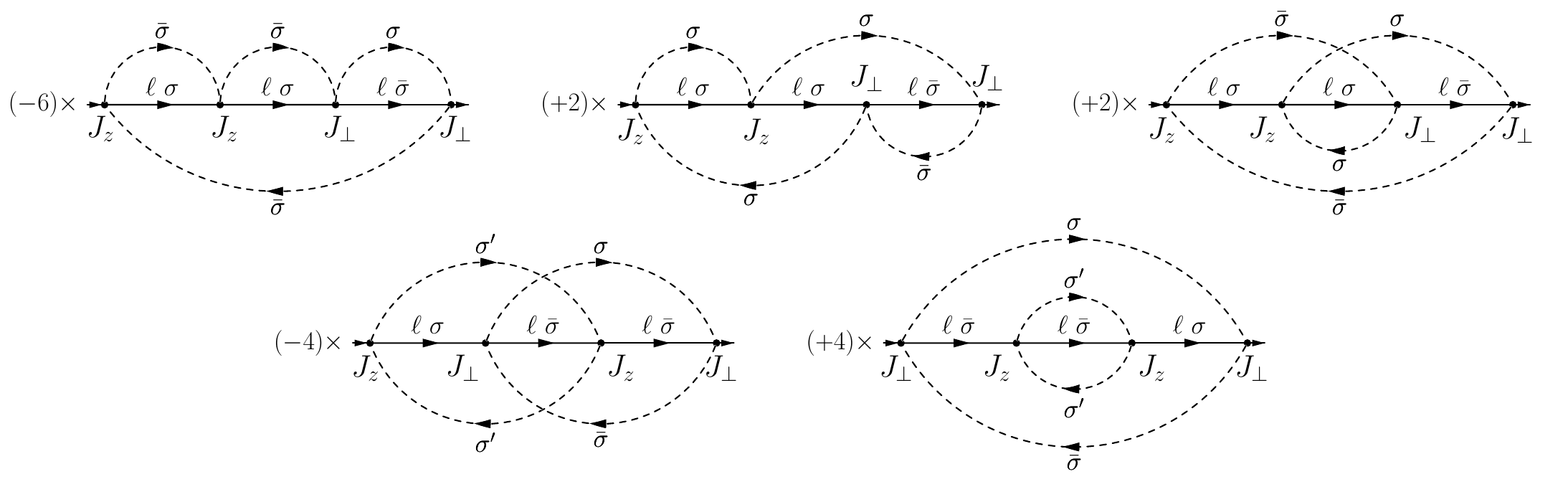}
\caption{Fourth-order diagrams contributing to the $J^{2}_{\perp} J_{z}^2$ term in the self-energy. The first row has one-pseudo\-fermion-loop diagrams while the second row has two-pseudo\-fermion-loop ones. Again, we show only hole diagrams, keeping in mind that the last two diagrams (second row) are symmetric to the reversal of pseudo\-fermion arrows and therefore do not have separate particle and hole versions.}
\label{fig:4thXloopSE2alt}
\end{figure*}

\begin{widetext}
Proceeding further, to the fourth order in the perturbative expansion, we encounter the diagrams shown in Fig.~\ref{fig:4thXloopSE}, which give a $J_{\perp}^{4}$ contribution to the self-energy. 
Their calculation proceeds in an analogous way as above
\footnote{At this order, the choice to do the calculation in $\tau$ space and Fourier transform the end result, instead of doing frequency summations, starts to show its advantages.},
and we find

\begin{equation}
\begin{split}
\Sigma_{(4,0) \perp} = - J_{\perp}^{4}  \rho_{0}^{3} \int_{-D}^{D} d \epsilon_{1} \int_{-D}^{D} d \epsilon_{2} \int_{-D}^{D} d \epsilon_{3} \bigg [ \frac{3(-2+\tanh(\frac{\beta \epsilon_2}{2})\tanh(\frac{\beta \epsilon_3}{2}))}{4(\epsilon_{1}-\epsilon_{2})(\epsilon_{1}-\epsilon_{3})(\epsilon_{1}-i \omega)} \bigg ]
\quad\mapsto\quad
\Sigma^{i}_{(4,0) \perp}=3 \pi  J_{\perp}^{4}   \rho_{0}^{3} \mathrm{ln}^{2}(\tilde{\omega})
\end{split}
\end{equation}
This is only a $J_{\perp}^{4}$ contribution to the self-energy; there is also an analogous $J_{\perp}^2 J_{z}^2$ contribution, with a larger number of diagrams as given in Fig.~\ref{fig:4thXloopSE2alt}. Adding up all those with their respective multiplicative factors we arrive at
\begin{equation}
\begin{split}
\Sigma_{(4,0) \perp z} = J_{\perp}^{2}J_{z}^{2}  \rho_{0}^{3} \int_{-D}^{D} d \epsilon_{1} \int_{-D}^{D} d \epsilon_{2} \int_{-D}^{D} d \epsilon_{3} \bigg [& \frac{3-4e^{\beta \epsilon_2}-4e^{\beta \epsilon_3}+3e^{\beta( \epsilon_2+\epsilon_3)}}{2(1+e^{\beta \epsilon_2})(1+e^{\beta \epsilon_3})(\epsilon_{1}-\epsilon_{2})(\epsilon_{1}-\epsilon_{3})(\epsilon_{1}-i \omega)}\\
\\
& + \frac{-1+\tanh(\frac{\beta \epsilon_2}{2})\tanh(\frac{\beta \epsilon_3}{2})}{2(\epsilon_{1}-\epsilon_{2})(\epsilon_{2}-\epsilon_{3})(\epsilon_{1}-i \omega)} \bigg ]
\quad\mapsto\quad
\Sigma^{i}_{(4,0)\perp z}=6 J_{\perp}^{2}J_{z}^{2}  \rho_{0}^{3}\mathrm{ln}^{2}(\tilde{\omega})
\end{split}
\end{equation}
\end{widetext}
(after analytic continuation, doing the energy integrals in the zero-temperature limit, and expanding in $\tilde{\omega}$).

Besides these two contributions, with $J_{\perp}^4$ and $J_{\perp}^2 J_{z}^2$, one can look for a $J_{z}^4$ contribution as well, but it turns out that this contribution is zero up to the dominant order we are keeping here.

\begin{figure}[t]
\includegraphics[width=0.49\textwidth]{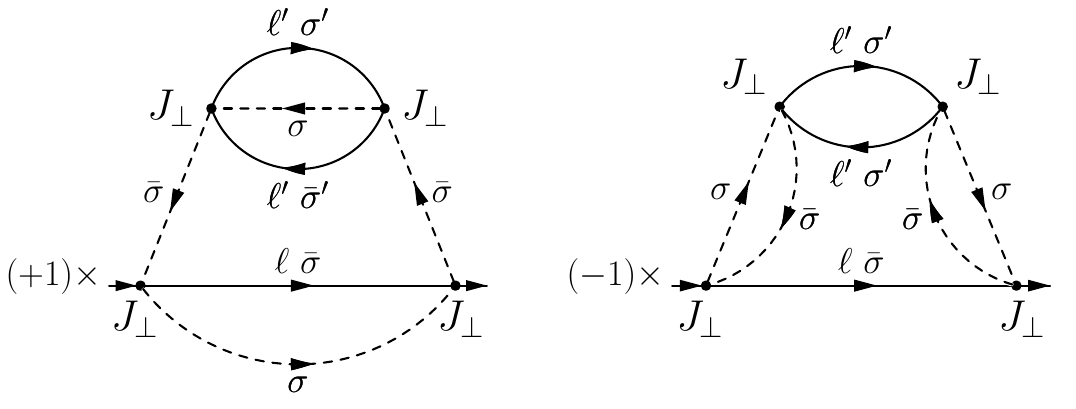}
\caption{Fourth-order diagrams contributing to the channel-number-dependent $J^{4}_{\perp}$ terms in the self-energy. The channel ``flavor'' in the fermionic loop is independent of the ``flavor'' of the fermions in the rest of the diagram (including also an $\alpha'$, not indicated). Therefore, the total channel number, $K=2M$, will appear multiplying their contribution to the self-energy (in addition to the multiplicative factors shown here).}
\label{fig:4thloopSE}
\end{figure}

There are additional fourth-order diagrams contributing to the electron self-energy that we did not touch upon as yet. Namely, these are the channel-number-dependent contributions, which, graphically, are given by diagrams with band-fermion loops. Some of such diagrams are shown in Fig.~\ref{fig:4thloopSE}, where we display the contributions with $J_{\perp}^4$. After the usual calculations, their dominant imaginary self-energy contribution is
\begin{equation}
\begin{split}
\Sigma^{i}_{(4,1) \perp} = 2J_{\perp}^{4} M  \pi \rho_{0}^{3}  \big [ \ln(2)-1 \big ] +  2 J_{\perp}^{4} M  \pi \rho_{0}^{3}  \ln(\tilde{\omega})
\end{split}
\end{equation}
where we have assumed an even number of channels and taken $K=2M$, with $M$ an integer (or half-integer) number.

\begin{figure}[b]
\includegraphics[width=0.49\textwidth]{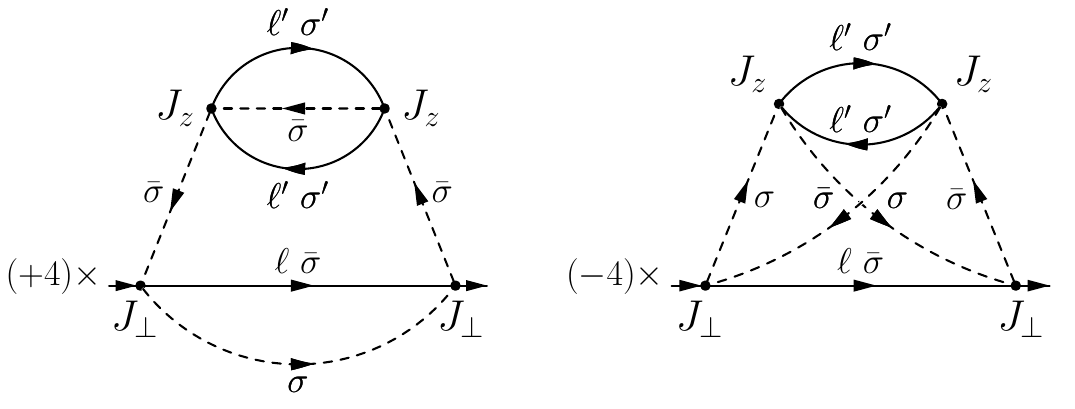}
\caption{Fourth-order diagrams contributing to the channel-number-dependent $J^{2}_{\perp} J_{z}^2$ terms in the self-energy. (Although it will not play a role in the current calculations because we are not doing impurity averaging, notice there is a single pseudo\-fermion loop in both diagrams, while in the previous figure the second diagram had two such loops.)}
\label{fig:4thloopSE2}
\end{figure}

Finally, we have also the diagrams in Fig.~\ref{fig:4thloopSE2}, which contribute to the $J_{\perp}^2 J_{z}^2$ terms in the electron self-energy. With a similar procedure as for all the previous cases we find
\begin{equation}
\begin{split}
\Sigma^{i}_{(4,1)\perp z}& = 4 J_{\perp}^{2}J_{z}^2  M  \pi \rho_{0}^{3}  \big [ \ln(2)-1 \big ]\\
& \qquad \qquad+4 J_{\perp}^{2}J_{z}^2 M  \pi \rho_{0}^{3}  \ln(\tilde{\omega})
\end{split}
\end{equation}
and, again, there are no $J_{z}^4$ contributions.

At this point, we have finished computing all the contributions to the perturbative expansion of the electron self-energy up to the fourth order in the coupling constant. Collecting all of the terms we arrive at the full expression for the imaginary part of the self energy. Writing it with increasing powers of the coupling (column\-wise) and decreasing degree of logarithmic divergence (row\-wise) one has

\begin{widetext}

\begin{alignat}{5}\label{eq:SEd}
\frac{4 \rho_{0}}{\pi}\Sigma^{i} = \ \  \bigg[P^{\perp}_{(2,0)} \mathrm{g}_{\perp}^{2} & \quad + \ P^{z}_{(2,0)} \mathrm{g}_{z}^{2}\bigg] \quad + & \ \bigg[P^{\perp z}_{(3,0)} \  \mathrm{g}_{\perp}^{2}\mathrm{g}_{z}\bigg] \mathrm{\ln} (\tilde{\omega}) & \quad + \ \bigg[P^{\perp}_{(4,0)} \  \mathrm{g}_{\perp}^{4}  & \quad & + \ P^{\perp z}_{(4,0)} \  \mathrm{g}_{\perp}^{2}\mathrm{g}_{z}^{2} \bigg] && \mathrm{ln}^{2} (\tilde{\omega})&  \nonumber \\
 & & & \quad + \ \bigg[P^{\perp}_{(4,1b)} \ M  \mathrm{g}_{\perp}^{4} & \quad &+ \ P^{\perp z}_{(4,1b)} \ M \mathrm{g}_{\perp}^{2}\mathrm{g}_{z}^{2}  \bigg] && \ln(\tilde{\omega})& \\
& & & \quad +  \ \bigg[ P^{\perp}_{(4,1a)} \ M \mathrm{g}_{\perp}^{4} & \quad &+ \ P^{\perp z}_{(4,1a)} \ M \mathrm{g}_{\perp}^{2}\mathrm{g}_{z}^{2}\bigg] &\nonumber
\end{alignat}
\vspace{0.7cm}
\end{widetext}

\noindent where the numerical coefficients in the right-hand side of the equation are

\begin{align}
P^{\perp}_{(2,0)} &= 2 , &
P^{z}_{(2,0)} &= 1 ,\nonumber\\
& & P^{\perp z}_{(3,0)} &= -12,\nonumber\\
P^{\perp}_{(4,0)} &= 12 , &
P^{\perp z}_{(4,0)} &= 24 , \label{eq:P3rdDirect} \\
P^{\perp}_{(4,1a)} &= 8 \big [ \ln(2)-1 \big ] ,&
P^{\perp z}_{(4,1a)} &= 16 \big [ \ln(2)-1 \big ] ,\nonumber\\
P^{\perp}_{(4,1b)} &= 8 , &
P^{\perp z}_{(4,1b)} &= 16 \nonumber
\end{align}

In Eq.~(\ref{eq:SEd}) we have made use of the substitution $\mathrm{g}=J\rho_{0}$, which is a usual convention (\textit{nota bene}~that latter we will redefine it as $g=2J\rho_{0}$, but we did not do it as yet in order to avoid for the time being fractional coefficients and connect with different conventions found in the literature; we shall always change the typography in order to alert the reader of this change taking place). Since the $P$ coefficients are numerical prefactors in front of different powers of the coupling constants, we have labeled them in a way that keeps track of which $P$ coefficient corresponds to which power. These coefficients are going to be used in determining the flow equations of the model. In general, we expect them to be different between the direct and conventional schemes (models). The reason being that in the conventional scheme we will have \textit{unphysical} diagrammatic contributions to the self-energy, which will change the numerical value of the $P$ coefficients. To what extent these changes actually reflect in the beta functions themselves will be discussed in detail in subsequent sections.

\subsection{Calculation of the flow}

As already argued above, considering that the imaginary part of the retarded self-energy corresponds to the scattering rate of the electrons by the impurity (a physical observable), it should be relatively insensitive to the cutoff scale, provided we allow the coupling constants to be adjusted. In other words, $\Sigma^{i}(\omega)$ should have a zero total derivative. This independence of the regulator scale is encapsulated in the CS equation [see Eq.~(\ref{eq:CS})], which can be rewritten in expanded form as
\begin{widetext}
\begin{equation}
\bigg ( -\frac{\partial}{\partial \mathrm{ln}(\tilde{\omega})}+\beta_{z} (\mathrm{g}_{z}, \mathrm{g}_{\perp})\frac{\partial}{\partial \mathrm{g}_{z}}+\beta_{\perp} (\mathrm{g}_{z}, \mathrm{g}_{\perp})\frac{\partial}{\partial \mathrm{g}_{\perp}} \bigg ) \ \Sigma^{i}(\tilde{\omega}, \mathrm{g}_{z}, \mathrm{g}_{\perp}) \ = \ 0
\label{eq:CS1}
\end{equation}
\end{widetext}
This expression is obtained by expanding the total derivative with respect to $\mathrm{\ln}(D)$ in terms of partial derivatives,
\begin{equation}
    \frac{d \Sigma^{i}}{d \mathrm{\ln}(D)} = \frac{\partial \Sigma^{i}}{\partial \mathrm{\ln}(D)} + \frac{\partial \mathrm{g}_{z}}{\partial \mathrm{\ln}(D)} \frac{\partial \Sigma^{i}}{\partial \mathrm{g}_{z}}+ \frac{\partial \mathrm{g}_{\perp}}{\partial \mathrm{\ln}(D)} \frac{\partial \Sigma^{i}}{\partial \mathrm{g}_{\perp}}
\end{equation}
replacing $\partial \Sigma^{i} / \partial \mathrm{\ln}(D) = - \partial \Sigma^{i} / \partial \mathrm{\ln}(\tilde{\omega})$, and introducing the definitions of the beta functions for the two coupling constants
\begin{equation}
\begin{split}
    &\beta_{z} (\mathrm{g}_{z}, \mathrm{g}_{\perp}) \equiv \frac{\partial \mathrm{g}_{z}}{ \partial \mathrm{\ln}(D)}\\
    &\beta_{\perp} (\mathrm{g}_{z}, \mathrm{g}_{\perp}) \equiv \frac{\partial \mathrm{g}_{\perp}}{ \partial \mathrm{\ln}(D)}
\end{split}
\end{equation}
From Eq.~(\ref{eq:CS1}), we can find the beta functions iteratively if we write them as a series expansion in the coupling constant with unknown coefficients (\textit{e.g.}, to third order)
\begin{equation}
\begin{split}\label{eq:beta0}
& \beta_{\perp}(\mathrm{g}_{z}, \mathrm{g}_{\perp}) =  a_{1} \mathrm{g}_{z}\mathrm{g}_{\perp}+ a_{2}  \mathrm{g}_{\perp}^3+ a_{3}\mathrm{g}_{\perp} \mathrm{g}_{z}^2\\
&  \beta_{z}(\mathrm{g}_{z}, \mathrm{g}_{\perp}) = b_{1} \mathrm{g}_{\perp}^2 + b_{2}\mathrm{g}_{z}^2+  b_{3}\mathrm{g}_{\perp}^2 \mathrm{g}_{z}+ b_{4} \mathrm{g}_{z}^3
\end{split}
\end{equation}
Since we know the first few terms of the perturbative expansion of $\Sigma^{i} (\tilde{\omega}, \mathrm{g}_{z}, \mathrm{g}_{\perp})$, we can perform all the derivatives in the CS equation to get a polynomial that can be arranged in powers of $\mathrm{g}_{z}$, $\mathrm{g}_{\perp}$, and $\mathrm{\ln} (\tilde{\omega})$, which are all regarded as independent. Notice we do not include in our power counting $M$ scaling as the inverse of the Kondo coupling, the way it is done in large-$M$ approaches \cite{gan1993,gan1994,bensimon2006}, but our conclusions will be nevertheless equivalent up to the third order 
\footnote{That modified power counting allows to ignore the fourth-order self-energy diagrams that contain no fermion-loops, but at this order those contribute only to additional consistency relations that change between schemes but are obeyed in both cases (see below).}. We can then equate with zero each combined factor in front of those powers and in such a way obtain a set of linear algebraic equations for the coefficients of the beta functions:
    \begin{align}
        a_{1} &= \frac{P_{(3,0)}^{\perp z}-2P_{(2,0)}^{z}b_{1}}{2 P_{(2,0)}^{\perp}}  & \quad b_{1} &= \frac{P_{(3,0)}^{\perp z} - 2 P_{(2,0)}^{\perp}a_{1}}{2 P_{(2,0)}^{z}}\nonumber\\
 a_{2} &= \frac{M P_{(4,1b)}^{\perp}}{2 P_{(2,0)}^{\perp}}  \ & \quad b_{2} &= 0 \label{eq:coeffD} \\
 a_{3} &=\frac{M P_{(4,1b)}^{\perp z}- 2 P_{(2,0)}^{z} b_{3}}{2 P_{(2,0)}^{\perp}} \ & \quad b_{3} &=\frac{M P_{(4,1b)}^{\perp z}-2 P_{(2,0)}^{\perp} a_{3}}{2 P_{(2,0)}^{z}} \nonumber\\
 & & \quad b_{4} &= 0 \nonumber
    \end{align}
Notice that the equations for $a_{1}$ and $b_{1}$, as well as those for $a_{3}$ and $b_{3}$, are the same equation just solved for a different variable. Thus we cannot solve simultaneously for $a_{1}$ and $b_{1}$ (respectively, $a_{3}$ and $b_{3}$), without introducing some other set of equations to determine them. In this case, an additional set of equations is easily obtained by considering the spin-isotropic limit. Namely, since we know that $\beta_{z}|_{\mathrm{g}_{z} = \mathrm{g}_{\perp} = \mathrm{g}} = \beta_{\perp}|_{\mathrm{g}_{z} = \mathrm{g}_{\perp} = \mathrm{g}} \equiv \beta(\mathrm{g})$, that enables us to extract two more equations that include the still unknown coefficients of the beta functions
\begin{equation}
\begin{split}
a_{1}&=b_{1}+b_{2}\\
a_{2}+a_{3}&=b_{3}+b_{4}
\end{split}
\end{equation}
These two additional relations help us solve the system and determine the beta functions to be
\begin{equation}
\begin{split}
\beta_{\perp}(\mathrm{g}_{z}, \mathrm{g}_{\perp}) &=  -2 \mathrm{g}_{z}\mathrm{g}_{\perp} + 2 M \left( \mathrm{g}_{\perp}^3+ \mathrm{g}_{\perp} \mathrm{g}_{z}^2 \right)\\
&\mapsto  -g_{z}g_{\perp} + \frac{M}{2} \left( g_{\perp}^3+ g_{\perp} g_{z}^2 \right)\\
\\
\beta_{z}(\mathrm{g}_{z}, \mathrm{g}_{\perp}) &= -2 \mathrm{g}_{\perp}^2+ 4M \mathrm{g}_{\perp}^2 \mathrm{g}_{z}\\
&\mapsto -g_{\perp}^2+ M g_{\perp}^2 g_{z}
\end{split}
\end{equation}
where, in the second lines, we introduced the rescaling $\mathrm{g}_{z,\perp}\mapsto g_{z,\perp}/2$ (and consequently $\beta_{z,\perp}\mapsto \beta_{z,\perp}/2$) to facilitate the comparisons with (our) poor man's scaling results (cf.~Ref.~\onlinecite{Hewson}) and other field-theoretical RG results in the literature that start with a different normalization of the initial Hamiltonian \cite{gan1994,bensimon2006}.

There are, however, additional equations to those given in Eq.~(\ref{eq:coeffD}), which we refer to as RG-consistency equations [since they indicate that the RG procedure can be used to ``consistently'' resum the logarithmic divergences row by row in Eq.~(\ref{eq:SEd})]. They are automatically satisfied by the solutions for the coefficients of the beta functions. Explicitly in this case, these equations are
\begin{equation}
    \begin{split}
     2 P_{(3,0)}^{\perp z} a_{1} & = 2 P_{(4,0)}^{\perp z} - P_{(3,0)}^{\perp z} b_{2}\\
     P_{(3,0)}^{\perp z} b_{1} & = 2 P_{(4,0)}^{\perp} \\
    \end{split}
    \label{eq:consistency}
\end{equation}
and are very interesting on more than one account. Firstly, they are the only equations that contain the coefficients of the $\mathrm{\ln^2}(\tilde{\omega})$ contributions to the electron self-energy. In fact, the system in Eq.~(\ref{eq:coeffD}) is derived fully from the $\mathrm{\ln}(\tilde{\omega})$ and nonlogarithmic contributions to the self-energy alone. Secondly, as one can also notice by inspecting the terms in Eq.~(\ref{eq:consistency}), they combine the lower-order coefficients of the beta function with the higher-order ones of the self-energy. As expected, the coefficients $a_{1}$ and $b_{1}$ can be completely derived using the self-energy up to the third order in the coupling constant, and one does not need to invoke higher-order terms of the self-energy to find them. This means that, from Eq.~(\ref{eq:consistency}), knowing $a_{1}$ and $b_{1,2}$ we can find the fourth-order $\mathrm{\ln^2}(\tilde{\omega})$ coefficients of the self-energy without actually doing any diagrammatic calculation (which we nevertheless did to find the less divergent parts). Notice that if the CS equation is found to be consistent in doing such a reverse calculation and producing the same result as the perturbative expansion, this constitutes a nontrivial check for evaluation of the higher-orders diagrams.

\section{RG calculation in the Indirect Schemes}
\vspace{0.4cm}
Having illustrated the field-theoretic RG procedure in detail for the (multi) two-channel Kondo model or the ``direct scheme'' we will move on to applying the same procedure for two different compactified versions of the model or ``indirect schemes''.
\vspace{0.1cm}
\subsection{BdB-compactified (multi) two-channel Kondo model}
The details of the BdB-based compactification of the (multi) two-channel Kondo model has been presented earlier \cite{Ljepoja2024a}. Here we simply recall the result of that procedure for the interaction part of the Hamiltonian,

\begin{equation}\label{eq::refermionized}
    \begin{split}
        H_{K} & = \tilde{n}_{c,\alpha}\tilde{n}^{-}_{l,\alpha}
        J_{\perp}\big(S^{-}\psi_{sl,\alpha}\psi^{\dagger}_{s,\alpha}-S^{+}\psi^{\dagger}_{sl,\alpha}\psi_{s,\alpha}\big)\\
        &\quad + \tilde{n}_{c,\alpha}\tilde{n}^{+}_{l,\alpha}
        J_{\perp}\big(S^{-}\psi_{sl,\alpha}^{\dagger}\psi_{s,\alpha}^{\dagger}-S^{+}\psi_{sl,\alpha}^{}\psi_{s,\alpha}^{}\big)\\
        &\quad + \tilde{n}_{c,\alpha}(\tilde{n}^{+}_{l,\alpha} +\tilde{n}^{-}_{l,\alpha})\, 
        J_{z} \ S^{z} : \! \psi^{\dagger}_{s,\alpha} \psi^{}_{s,\alpha} \! : \\
        &\quad + \tilde{n}_{c,\alpha} (\tilde{n}^{+}_{l,\alpha} -\tilde{n}^{-}_{l,\alpha})\,
        J_{z} \ S^{z} : \! \psi^{\dagger}_{sl,\alpha}\psi_{sl,\alpha}\! :
    \end{split}
\end{equation}

\noindent where the fermionic fields are evaluated at the position of the impurity ($x\!=\!0$). Succinctly, the expression in Eq.~(\ref{eq::refermionized}) is obtained by carrying out the standard bosonization procedure, after which we rotate the model, for each value of $\alpha$, into so-called \textit{physical sectors} where (instead of two-valued ``spin'' and ``lead'' degrees of freedom) we have $c$, $s$, $l$, and $sl$ labels \cite{Ljepoja2024a}, and finally we end by de\-bosonizing. Via such rotations, the original model is mapped into one where only two sectors would be coupled to the impurity, namely, $s$ and $sl$, as expected on physical grounds since the impurity is a pure-spin degree of freedom. However, we have argued that one needs to take a departure from the ``conventional'' way of handling these transformations, and introduce the $\tilde{n}$ factors whereby the sectors that are not directly coupled by the impurity dynamics are still correlated to it. Intuitively, these factors can be thought of as fermionic densities, but their rigorous origin is in the products of (fractional) vertex operators of the respective bosonic fields. The standard, what we refer to as \textit{conventional}, compactification result is recovered by letting all $\tilde{n}\to 1$. This limit, however, can produce unphysical results, since it allows for processes which have no counterpart in the language of the original fermions. To avoid that, we introduced the \textit{consistent} scheme, in which the $\tilde{n}$ factors are preserved as adiabatic operators and one utilizes their inherent properties of idempotence ($\tilde{n}^2\!=\!\tilde{n}$) and co-nilpotence ($\tilde{n}^{+} \tilde{n}^{-}\!=\!0$) for each fermion history \cite{Ljepoja2024b}. We shall now compare these two schemes between them and with the direct one (since they can all be thought of as different schemes for determining the universal RG flow of the model).

\begin{widetext}
\subsection{Conventional Scheme}

The conventionally refermionized action has a standard Gaussian part (including pseudofermions),
\begin{equation}
S_{0} =  \rho_{0} \sum_{\nu,\alpha}\int_{-D}^{D} d\epsilon \int_{0}^{\beta} d \tau \bar{\psi}_{\nu \alpha}(\epsilon, \tau)(\epsilon +\partial_{\tau}) \psi_{\nu \alpha}(\epsilon, \tau)
+ \sum_{\sigma} \int_{0}^{\beta} d \tau \bar{\eta}_{\sigma}(\tau) (\mu + \partial_{\tau})  \eta_{\sigma}(\tau)\\
\end{equation}
where $\nu\!=\!\{c, s, l, sl \}$ labels the physical sector, and $\alpha\!=\!1,\dots,M$ labels the (additional) channel degeneracy for each sector. On the other hand, the impurity interaction part of the action is given by

\begin{equation}
\begin{split}
S_{I}  = & -J_{+} \rho_{0}^2 \sum_{\alpha}\int_{-D}^{D} \ d \epsilon\int_{-D}^{D} \ d \epsilon^{\prime}\int_{0}^{\beta} d \tau \ \bar{\eta}_{\downarrow} (\tau) \eta_{\uparrow}(\tau) \bar{\psi}_{s , \alpha}(\epsilon, \tau) \psi_{sl , \alpha}(\epsilon^{\prime}, \tau) \\
& - J_{-}  \rho_{0}^2 \sum_{\alpha}\int_{-D}^{D} \ d \epsilon\int_{-D}^{D} \ d \epsilon^{\prime}\int_{0}^{\beta} d \tau \ \bar{\eta}_{\uparrow} (\tau) \eta_{\downarrow}(\tau) \bar{\psi}_{sl , \alpha}(\epsilon, \tau)\psi_{s , \alpha}(\epsilon^{\prime}, \tau)\\
& - J^{A}_{+}  \rho_{0}^2 \sum_{\alpha}\int_{-D}^{D} \ d \epsilon\int_{-D}^{D} \ d \epsilon^{\prime}\int_{0}^{\beta} d \tau \ \bar{\eta}_{\uparrow} (\tau) \eta_{\downarrow}(\tau) \psi_{sl, \alpha}(\epsilon, \tau)\psi_{s , \alpha}(\epsilon^{\prime}, \tau)\\
& - J^{A}_{-}  \rho_{0}^2 \sum_{\alpha}\int_{-D}^{D} \ d \epsilon\int_{-D}^{D} \ d \epsilon^{\prime}\int_{0}^{\beta} d \tau \ \bar{\eta}_{\downarrow} (\tau) \eta_{\uparrow}(\tau)\bar{\psi}_{s , \alpha}(\epsilon, \tau) \bar{\psi}_{sl , \alpha}(\epsilon^{\prime}, \tau)\\
& + J_{z}  \rho_{0}^2 \sum_{\alpha} \int_{-D}^{D} \ d \epsilon\int_{-D}^{D} \ d \epsilon^{\prime}\int_{0}^{\beta} d \tau \ \bigg ( \bar{\eta}_{\uparrow} (\tau) \eta_{\uparrow}(\tau)  - \bar{\eta}_{\downarrow} (\tau) \eta_{\downarrow}(\tau) \bigg  ) \bar{\psi}_{s , \alpha}(\epsilon, \tau)\psi_{s , \alpha}(\epsilon^{\prime}, \tau)
\end{split}
\end{equation}
\end{widetext}

\noindent where we have introduced besides $J_{\sigma}$ also $J_{\sigma}^{A}$ as coupling constant notations just to highlight the terms that involve \textit{anomalous} vertices ($A$ superscript). The value of $\sigma\!=\!\pm$ indicates whether the impurity is flipping its spin up or down, respectively. This notation is introduced just for the sake of easier reading of the diagrams and we still maintain, for calculations purposes, $J_{\sigma}^{A}\!=\!J_{\sigma}\!=\!J_{\perp}$, for both values of $\sigma$. Tracking the two types of vertices is interesting because it is the mixing of \textit{anomalous} vertices with the \textit{normal} ones that introduces unphysical contributions, and all of those come initially from the diagrams contributing to $J_{\perp}^4$ in the self-energy. There are two important differences in terms of the ``diagrammar'' of the conventional scheme versus the direct one: first, the presence of the anomalous terms will produce diagrams with reversed arrows on the fermion propagators and, second, if one looks at the $J_{z}$ vertex one notices that it involves only $s$ fermion scattering (the $sl$ sector does not participate). In turn, this means that one cannot put a $J_{z}$ vertex in between two $J_{\perp}$ ones, which limits the number of diagrams that can be made.

\begin{figure}[b!]
\includegraphics[width=0.49\textwidth]{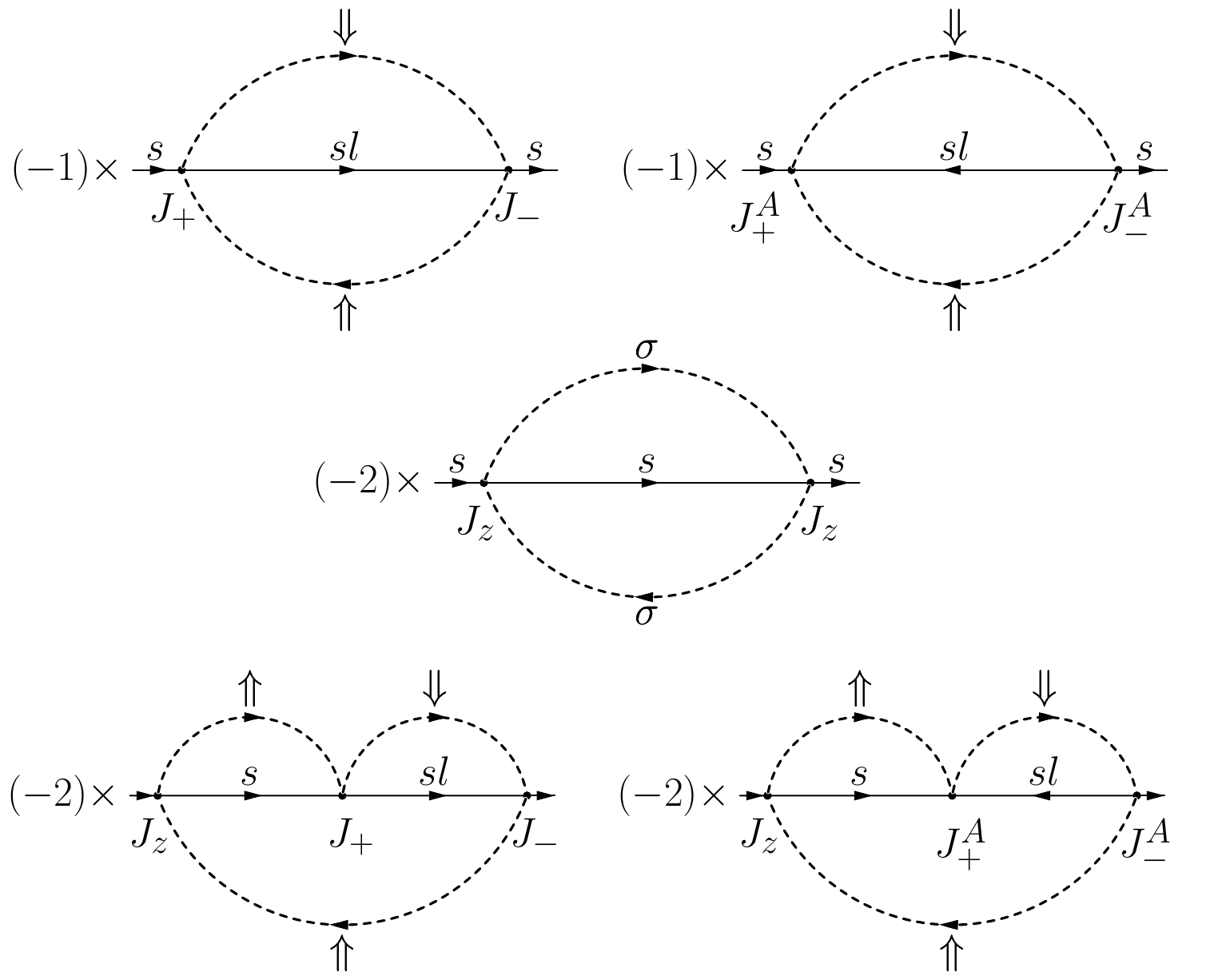}
\caption{Second- and third-order diagrams contributing to the $s$-fermion self-energy (for a given $\alpha$). At these orders of perturbation theory there are no unphysical contributions present. For the second-order diagrams we labeled the external ``amputated'' legs, (as we did also in Fig.~\ref{fig:2nd3rdSE}), but we dropped those labels in higher-order diagrams.}
\label{fig:2nd3rdSEc}
\end{figure}

It is interesting to make an aside and point out how corresponding conventional- and direct-scheme diagrams are related. The reversal of propagators in the diagrams can be accounted for by observing that the Green functions obey: $G^{f}_{0}(-\tau,\epsilon) = -G^{f}_{0}(\tau,-\epsilon) = -G^{f}_{0}(\tau, \epsilon)$; where the last equality is valid only under the integral in $\epsilon$ with symmetric integration limits. This way any propagator oriented ``leftward'' in a conventional-scheme diagram can be made into a ``rightward'' one as in the direct scheme. In the process, it incurs an extra minus sign which is compensated by another minus sign coming from the Wick contractions of anomalous vertices. In Fig.~\ref{fig:2nd3rdSEc} we show the diagrams that contribute at the second and third orders in the coupling constant to the $s$-fermion self-energy. Notice that in the diagrams with anomalous vertices the $sl$-fermion propagator appears ``reversed''. Using the property mentioned above the arrow can be ``turn back right'', thereby making those diagrams the same as the others at the same respective order; but that introduces a minus sign. However, the Wick contraction of two anomalous vertices also produces an extra minus sign (as opposed to the contraction of normal ones). As a result, all the diagrams in the figure that have fermion propagators going leftward end up giving the same contributions as the corresponding rightward diagrams.

Explicitly, every second-order diagram in 
Fig.~\ref{fig:2nd3rdSEc} gives the same contribution, and collecting them we arrive at
\begin{equation}
\begin{split}
&\Sigma^{s, i}_{(2,0) \perp} = \pi \rho_{0}  J_{\perp}^{2} \\
&\Sigma^{s, i}_{(2,0) z} = \pi   \rho_{0}  J^{2}_{z}
\end{split}
\end{equation}
and the same is true for the third-order diagrams that evaluate to the same result and give a combined contribution to the imaginary part of the self-energy given by
\begin{equation}
\begin{split}
\Sigma^{s,i}_{(3,0) \perp z} = -8 \pi \rho_{0}^{2} J_{\perp}^2 J_{z}   \ln(\tilde{\omega})
\end{split}
\end{equation}

Even though in Fig.~\ref{fig:2nd3rdSEc} we have diagrams that have anomalous vertices, these are not unphysical diagrams. They are perfectly translatable in terms of direct-scheme fermions, because they do not mix normal and anomalous $J_{\perp}$ vertices. They correspond, in principle, to different lead contributions. Diagrams with all normal vertices will translate into contributions to the $L$-electron self-energy while the ones with all anomalous vertices will do the same for the $R$-electron self-energy. Since the problems with the translatability of conventional-scheme diagrams arise from diagrams that mix normal and anomalous vertices, any diagrams that involve only two spin-flip vertices are not able to produce the said mixing in the self-energy expansion; remark that the external legs of the diagrams (the ones that are \textit{amputated}) have to be oriented in the same direction. In other words, if in the second-order diagram we mixed a normal and an anomalous vertex, it would produce a diagram that has both external $s$-fermion propagators going into the diagram, which is not a self-energy diagram. This means that we expect the leading-order beta functions to be the same between conventional and direct schemes, and that for the discrepancy between the two schemes to arise one should look at higher orders of the coupling-constant expansion. 

\begin{figure}[t!]
\includegraphics[width=0.49\textwidth]{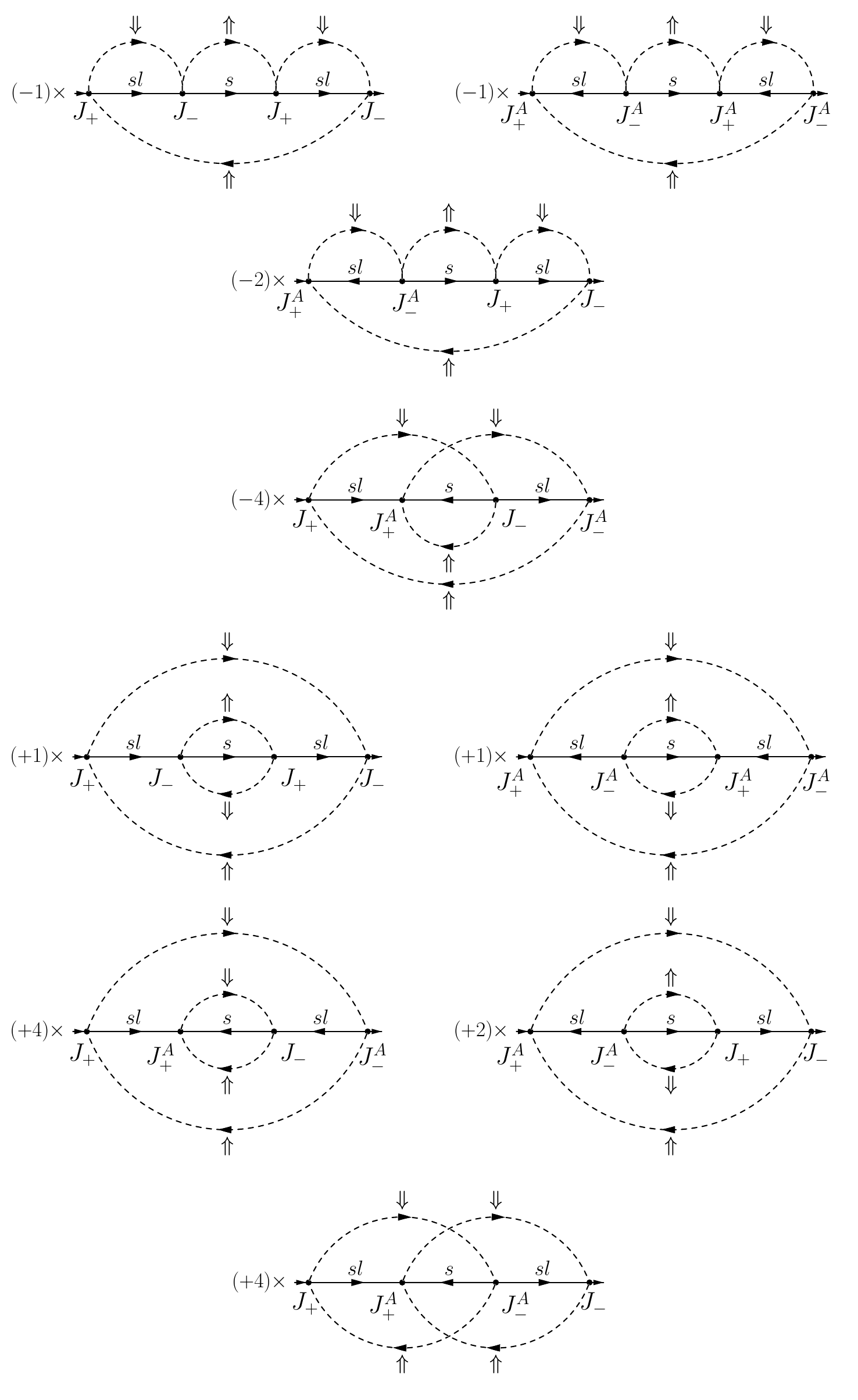}
\caption{Fourth-order diagrams contributing to the $J^{4}_{\perp}$ term in the conventional self-energy. Unphysical diagrams are those that ``mix'' normal and anomalous vertices. Such diagrams are shown in rows two and three (with a single pseudo\-fermion loop), as well as five and six (with two pseudo\-fermion loops). These contributions will not be possible to translate them into the language of the original $L$ and $R$ fermions.}
\label{fig:4thXloopSEc}
\end{figure}

At the fourth order in $J$, there is only one type of contribution in which unphysical diagrams can appear, namely, the $J_{\perp}^4$ contribution. Also, in addition to having these unphysical contributions from diagrams similar to the ones we had before, there will be additional diagrammatic contributions to this order in the self-energy that were not present in the direct-scheme calculations. All of this comes across precisely in Fig.~\ref{fig:4thXloopSEc}, where only diagrams in the first and fourth rows are not unphysical. Comparing this set of diagrams with the one that contributes to $J_{\perp}^4$ in the direct scheme (cf.~Fig.~\ref{fig:4thXloopSE}), one can see that (in the conventional scheme) besides the unphysical versions of the direct-scheme diagrams, one also encounters ``topologically'' new diagrammatic contributions. The latter are shown in rows three and six in the figure, and both of them are unphysical. The reason they are disallowed in the direct scheme is because they have two $J_{+}$ (respectively $J_{-}$) vertices next to each other, which in the direct scheme would mean to flip the fermion spin up (respectively down) two times in sequence. There are no such contributions in the direct scheme, for a spin-$1/2$ can be flipped only once consecutively in each direction. All the above considerations about physicality notwithstanding, the contribution from the $J_{\perp}^4$ diagrams is
\begin{equation}
\Sigma^{s,i}_{(4,0) \perp} = 8 \pi J_{\perp}^{4} \rho_{0}^{3} \ln^{2}(\tilde{\omega}) -\frac{\pi^3}{4} J_{\perp}^{4} \rho_{0}^{3}
\end{equation}
Comparing it to the direct-scheme contribution, we see that in addition to the coefficient in front of $\mathrm{\ln}(\tilde{\omega})$ being different, there is a constant term (\textit{i.e.}, independent of $\tilde{\omega}$) present as well. The appearance of this constant term will not contribute to the beta function at this order (\textit{i.e.}, third order in the coupling). 
We leave aside for the moment the possibility of a modified beta function at higher orders (which requires the inclusion of fifth- and higher-order contributions to the self-energy).
On the other hand, the fact that the $P^{\perp}_{(4,0)}$ coefficient in the conventional-scheme self-energy is different than in the direct scheme, affects only the RG-consistency equation [cf.~Eq.~(\ref{eq:consistency})] and not the actual equations for the coefficients of the beta functions given in Eq.~(\ref{eq:coeffD}). Since one expects the conventionally refermionized theory to be RG-consistent by itself, one expects that this modified conditions will be satisfied with the modified coefficients of the fourth-order self-energy. In the end, we expect the beta functions to be the same between both schemes at this order, since nowhere else besides the RG-consistency conditions do the unphysical diagrams contribute.

The other contribution to the fourth-order self-energy is the $J^2_{\perp} J^2_{z}$ one. This one, however, does not include any unphysical diagrams, since there is no way one can mix normal and anomalous spin-flip vertices because there are only two of those per diagram. (Each diagram can thus only have either all normal or all anomalous spin-flip vertices.) So, in principle, we shall have the same diagrams as in Fig.~\ref{fig:4thXloopSE2alt} except that they come with different multiplicative factors. This makes the contribution of the $J^{2}_{\perp} J^2_{z}$ term be
\begin{equation}
\begin{split}
\Sigma^{s,i}_{(4,0) \perp z} = 16 \pi J_{\perp}^{2}J_{z}^{2} \rho_{0}^{3}  \ln^{2}(\tilde{\omega})+\frac{\pi^4}{4} J_{\perp}^{2}J_{z}^{2} \rho_{0}^{3}
\end{split}
\end{equation}
As in the case of the $J_{\perp}^4$ contribution, we have an additional constant term, which, up to the current order of calculation, does not affect the beta function.

As for as the channel-number-dependent contributions (the ones coming from diagrams with fermion loops), they do not include any unphysical contribution. The reason being the same as for the second-order contributions. If one looks closely at the $J_{\perp}^4$ fermion-loop diagrams (cf.~Fig.~\ref{fig:4thloopSE}), we see that the lower two vertices are similar as in the second order diagrams. So they both have to be either normal or anomalous, and mixing is not possible. Furthermore, the upper two vertices involved in the fermionic loop are separated from the bottom two, and they can also both be anomalous or normal (independent of the bottom ones), and mixing is again not possible. In particular, there could be a diagram that has two anomalous vertices in the fermionic loop, while the other two vertices are normal (or \textit{vice versa}). That, however, would be perfectly translatable to the direct scheme and not unphysical. The $\alpha$ index in the fermion loop is independent of that in the base line of the diagram and contributes an $M$ multiplicity.
Thus, the conventional-scheme contribution from fermion-loop diagrams is
\small
\begin{equation}
\begin{split}
& \Sigma^{s,i}_{(4,1)\perp} = 4  \pi J_{\perp}^{4} M \rho_{0}^{3}  \big [ \ln(2)-1 \big ]+4  \pi J_{\perp}^{4} M \rho_{0}^{3}  \ln(\tilde{\omega})\\
\\
& \Sigma^{s,i}_{(4,1)\perp z} = 12 \pi J_{\perp}^{2}J_{z}^{2} M \rho_{0}^{3}  \big [ \ln(2)-1 \big ]+ 12  \pi J_{\perp}^{2}J_{z}^2 M \rho_{0}^{3}  \ln(\tilde{\omega})
\end{split}
\end{equation}
\normalsize

Once again we find a slightly different result as compared to the direct scheme. This is a consequence of the $J_{z}$ vertex scattering involving only the $s$-sector fermions and having physically allowed mixing of normal and anomalous vertices. It is interesting to note that this allowed mixing of normal and anomalous vertices in the conventional scheme is analogous to the mixing of different channels in the same diagrams in the direct scheme.

\begin{widetext}
Collecting all the contributions, as we did in the direct-scheme calculation, we arrive at
\begin{alignat}{5}
\frac{4\rho_{0}}{\pi}\Sigma^{i}_{s} = \ \  \bigg[P^{\perp}_{(2,0)} \mathrm{g}_{\perp}^{2} & \quad + \ P^{z}_{(2,0)} \mathrm{g}_{z}^{2}\bigg] \quad + & \ \bigg[P^{\perp z}_{(3,0)} \  \mathrm{g}_{\perp}^{2}\mathrm{g}_{z}\bigg] \mathrm{\ln} (\tilde{\omega}) & \quad + \ \bigg[P^{\perp}_{(4,0)} \  \mathrm{g}_{\perp}^{4}  & \quad & + \ P^{\perp z}_{(4,0)} \  \mathrm{g}_{\perp}^{2}\mathrm{g}_{z}^{2} \bigg] && \mathrm{ln}^{2} (\tilde{\omega})&  \nonumber \\
 & & & \quad + \ \bigg[P^{\perp}_{(4,1b)} \ M  \mathrm{g}_{\perp}^{4} & \quad &+ \ P^{\perp z}_{(4,1b)} \ M \mathrm{g}_{\perp}^{2}\mathrm{g}_{z}^{2}  \bigg] && \ln(\tilde{\omega})& \nonumber \\
& & & \quad +  \ \bigg[ P^{\perp}_{(4,0a)}  \mathrm{g}_{\perp}^{4} & \quad &+ \ P^{\perp z}_{(4,0a)} \mathrm{g}_{\perp}^{2}\mathrm{g}_{z}^{2} \bigg] &\nonumber \\
& & & \quad +  \ \bigg[ P^{\perp}_{(4,1a)} \ M \mathrm{g}_{\perp}^{4} & \quad &+ \ P^{\perp z}_{(4,1a)} \ M \mathrm{g}_{\perp}^{2}\mathrm{g}_{z}^{2} \bigg] &
\label{eq:SEc}
\end{alignat}
\end{widetext}
\noindent where the \textit{conventionally} calculated coefficients are now given by

\begin{align}
&P^{\perp}_{(2,0)} = 4  &  &P^{z}_{(2,0)} = 4 \nonumber\\
&&  &P^{\perp z}_{(3,0)} = -32\nonumber\\
&P^{\perp}_{(4,0)} = 32  &  &P^{\perp z}_{(4,0)} = 64 \label{eq:P3rdConventional} \\
&P^{\perp}_{(4,1a)} = 16 \big [ \ln(2)-1 \big ] & \quad &P^{\perp z}_{(4,1a)} = 48 \big [ \ln(2)-1 \big ] \nonumber\\
&P^{\perp}_{(4,1b)} = 16   &  &P^{\perp z}_{(4,1b)} = 48 \nonumber\\
&P^{\perp}_{(4,0a)} = -\pi^3   &  &P^{\perp z}_{(4,0a)} = \pi^3 \nonumber
\end{align}

It is evident that in the conventional scheme we have coefficients that are different than in the direct calculation. The question is whether this will lead to different beta functions. The ansatz for the beta functions and the set of equations to determine their coefficients turn out to be the same as they were in the direct scheme and given by Eq.~(\ref{eq:beta0}) and Eq.~(\ref{eq:coeffD}), respectively. Filling in with the now-different self-energy $P$ coefficients of Eq.~(\ref{eq:P3rdConventional}), one nevertheless arrives at exactly the same beta functions as in the direct case (with all of the RG-consistency equations satisfied as well). The additional $\tilde{\omega}$-independent terms that appear in the conventional scheme, as expected, do not change the third-order beta functions. Remarkably, the presence of the unphysical diagrams does not affect the universal physics of the model (see our conclusions from poor man's scaling \cite{Ljepoja2024b}). 

\begin{widetext}
\subsection{Consistent scheme}

The consistently refermionized action is given by the same Gaussian terms as above and the following Kondo-interaction terms:
\begin{equation}
\begin{split}
S_{I} & = -\tilde{n}_{c,\alpha} \tilde{n}^{-}_{l,\alpha} J_{-}\int_{-D}^{D} d \epsilon \int_{-D}^{D} d \epsilon^{\prime} \int_{0}^{\beta} d \tau \bar{\eta}_{\downarrow} (\tau) \eta_{\uparrow}(\tau) \bar{\psi}_{s,\alpha}(\tau, \epsilon) \psi_{sl,\alpha}(\tau, \epsilon^{\prime})\\
& \quad -  \tilde{n}_{c,\alpha} \tilde{n}^{-}_{l,\alpha} J_{+}\int_{-D}^{D} d \epsilon \int_{-D}^{D} d \epsilon^{\prime} \int_{0}^{\beta} d \tau \bar{\eta}_{\uparrow} (\tau) \eta_{\downarrow}(\tau) \bar{\psi}_{sl,\alpha}(\tau, \epsilon)\psi_{s,\alpha}(\tau, \epsilon^{\prime})\\
&\quad -  \tilde{n}_{c,\alpha} \tilde{n}^{+}_{l,\alpha} J^{A}_{-}\int_{-D}^{D} d \epsilon \int_{-D}^{D} d \epsilon^{\prime} \int_{0}^{\beta} d \tau \bar{\eta}_{\downarrow} (\tau) \eta_{\uparrow}(\tau)\bar{\psi}_{s,\alpha}(\tau, \epsilon^{\prime}) \bar{\psi}_{sl,\alpha}(\tau, \epsilon)\\
& \quad -  \tilde{n}_{c,\alpha} \tilde{n}^{+}_{l,\alpha} J^{A}_{+}\int_{-D}^{D} d \epsilon \int_{-D}^{D} d \epsilon^{\prime} \int_{0}^{\beta} d \tau \bar{\eta}_{\uparrow} (\tau) \eta_{\downarrow}(\tau) \psi_{sl,\alpha}(\tau, \epsilon)\psi_{s,\alpha}(\tau, \epsilon^{\prime})\\
&  \quad + \tilde{n}_{c,\alpha} \tilde{n}^{-}_{l,\alpha} J_{z} \int_{-D}^{D} d \epsilon \int_{-D}^{D} d \epsilon^{\prime} \int_{0}^{\beta} d \tau \big ( \bar{\eta}_{\uparrow} (\tau) \eta_{\uparrow}(\tau)  - \bar{\eta}_{\downarrow} (\tau) \eta_{\downarrow}(\tau) \big  ) \big (\bar{\psi}_{s,\alpha}(\tau, \epsilon)\psi_{s,\alpha}(\tau, \epsilon^{\prime})-\bar{\psi}_{sl,\alpha}(\tau, \epsilon)\psi_{sl,\alpha}(\tau, \epsilon^{\prime}) \big )\\
&  \quad + \tilde{n}_{c,\alpha} \tilde{n}^{+}_{l,\alpha} J_{z} \int_{-D}^{D} d \epsilon \int_{-D}^{D} d \epsilon^{\prime} \int_{0}^{\beta} d \tau \big ( \bar{\eta}_{\uparrow} (\tau) \eta_{\uparrow}(\tau)  - \bar{\eta}_{\downarrow} (\tau) \eta_{\downarrow}(\tau) \big ) \big (\bar{\psi}_{s,\alpha}(\tau, \epsilon)\psi_{s,\alpha}(\tau, \epsilon^{\prime}) + \bar{\psi}_{sl,\alpha}(\tau, \epsilon)\psi_{sl,\alpha}(\tau, \epsilon^{\prime}) \big )
\end{split}
\end{equation}
\end{widetext}
One can see that, besides the $\tilde{n}$ factors that serve to prevent the appearance of unphysical diagrams, the biggest difference between the consistent and conventional compactifications is that the $J_{z}$ vertex now scatters both $s$- and $sl$-sector fermions. Since the two sectors are now contributing in a more symmetric way to the action, we expect that the $s$ and $sl$ self-energies will be at a similar footing. This is already quite a contrast to the conventional scheme, where using the $sl$ self-energy leads to a smaller number of algebraic constraints and is not enough to determine the coefficients of the beta functions to the desired order (see Appendix~\ref{Apdx:slSelfEnergy} for more details).

Having the $\tilde{n}$ consistency factors present, as already explained, prevents the mixing of anomalous and normal vertices in any given fermion line of the diagrams. (Some mixing is thus allowed in fermion-loop diagrams, since it does not produce unphysical contributions.)
The diagrams then look the same as in the direct case, except for some of the arrow orientations and the presence of the $\tilde{n}_{l,\alpha}^{\mp}$ factors at the vertices. They correspond to all-normal and all-anomalous vertices (per fermion line), respectively. There are no contributions from mixing normal and anomalous spin-flip processes, since such mixing is prevented by the co-nilpotence of the consistency factors. The multiplicative factors in front of the integrals are thus the same as in the direct scheme, thereby producing the same self-energy result as in that scheme for each of the $\tilde{n}$ contributions. To put it more precisely, the result for the consistent-scheme imaginary-part of the retarded self-energy is related to the direct-scheme one by
\begin{equation}
    \begin{split}
        \Sigma^{i}_\text{Consistent} = ( \tilde{n}_{l,\alpha}^{-} + \tilde{n}_{l,\alpha}^{+} ) \Sigma^{i}_\text{Direct} = \Sigma^{i}_\text{Direct}
    \end{split}
    \label{eq:consSE}
\end{equation}
where we have used the properties of idempotence and ``completeness'' ($\tilde{n}_{l,\alpha}^{-} + \tilde{n}_{l,\alpha}^{+}=1$). This means that the beta functions produced from the consistent-scheme calculation are the same as in the direct scheme and thus coincide also (at this order) with the conventional one as well. Regardless of the matching results, the advantages of the consistent scheme over the conventional one are that we can use either the $s$ or $sl$ self-energy to find the beta functions, and we avoid any unphysical contributions in a systematic way.

Another way one can be convinced of the fact that the consistent and direct schemes produce the same results (at all orders) is by noticing that the set of (mapping) transformations
\begin{equation}
    \begin{split}
        \tilde{n}_{l, \alpha}^{-} \bar{\psi}_{s, \alpha} \quad &\mapsto \quad \bar{\psi}_{\uparrow L \alpha}\\
        \tilde{n}_{l, \alpha}^{-} \bar{\psi}_{sl, \alpha} \quad &\mapsto \quad -\bar{\psi}_{\downarrow L \alpha}\\
        \tilde{n}_{l, \alpha}^{+} \bar{\psi}_{s, \alpha} \quad &\mapsto \quad \bar{\psi}_{\uparrow R \alpha}\\
        \tilde{n}_{l, \alpha}^{-} \bar{\psi}_{sl, \alpha} \quad &\mapsto \quad - \psi_{\downarrow R \alpha}
    \end{split}
\end{equation}
returns the consistently refermionized action back into the (original) direct one. Therefore, it is to be expected that they will produce identical results in all respects. One can think of the consistently compactified model as a rewrite of the original model in terms of physical-sector degrees of freedom. 

\subsection{Comparison of results up to third order}

We found that all three schemes for calculating the RG flow of the (multi) two-channel Kondo model give the same result for the third-order beta functions, namely,
\begin{equation}
\begin{split}
\beta_{\perp}(g_{z}, g_{\perp}) &= -g_{z}g_{\perp} + \frac{M}{2} \left( g_{\perp}^3+ g_{\perp} g_{z}^2 \right)\\
\\
\beta_{z}(g_{z}, g_{\perp}) &= -g_{\perp}^2+ M g_{\perp}^2 g_{z}
\end{split}
\end{equation}

This is the same result one obtains by doing poor man's scaling \cite{Ljepoja2024b}. This means that all three schemes capture well the universal physics of the Kondo model. This is a not unexpected though somewhat surprising result, considering the appearance of some unphysical processes in the conventional scheme. However, as we have seen, those contributions end up affecting only the RG-consistency equations and not the equations used to calculate the coefficients of the beta functions. Nevertheless, the self-energy (and therefore the electron scattering rate) after the conventional compactification is different than in the other two models. So, one should expect differences among them to manifest away from the IR fixed point; \textit{i.e.}, at relatively high temperatures or in nonequilibrium quantities (like the differential spin conductance that we discussed earlier \cite{Ljepoja2024a}).

\section{RG flow beyond third order} \label{Sec:Beyond}

So far we have calculated the RG flow up to the third order in the coupling constant. We have seen that although there are, in the conventional scheme, unphysical terms contributing to the imaginary part of the retarded self-energy, the flow remains the same as it was in the direct calculation. 
Comparing this to our previous results \cite{Ljepoja2024a,Ljepoja2024b}, we observe that it fits into a general narrative that despite certain physical quantities being different between schemes, 
the universal physics remains the same. 

\begin{figure}[h]
\includegraphics[width=0.49\textwidth]{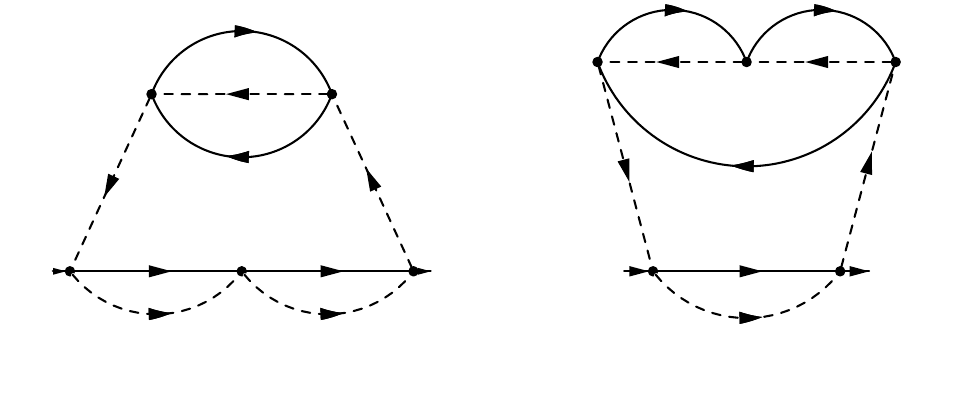}
\caption{Examples of fifth-order diagrams with fermionic loops. Each of the two diagrams will have many more pseudo\-fermionic contractions than shown, (but for the analysis of physicality only the full-line fermionic propagators are needed to check for mixing of normal and anomalous vertices). These diagrams, with the fermionic contractions as shown, do not harbor any unphysical contributions.}
\label{fig:5thloopSE}
\end{figure}

In the next few subsections, we would like to compare the schemes going beyond the third-order flow. More precisely, we want to calculate the channel-number-independent fourth-order terms in the beta functions. We want to check whether at this order (which will bring in more egregious examples of unphysical contributions), the beta functions are still the same between schemes or not. In order to do such a comparison, we compute the self-energy up to the fifth order in the coupling constant but taking into account only diagrams that have no fermion loops. Calculations of high-order fermion-loop diagrams can be found elsewhere \cite{gan1994}, and they do not involve any unphysical contributions when in the conventional scheme; so for our current purposes of comparison they are of little value. Schematic examples of such diagrams are shown in Fig.~\ref{fig:5thloopSE}, and from there it is obvious why they will purport no unphysical contributions: the bottom part of the diagram is effectively either a third- or second-order diagram, respectively, and as such cannot mix normal and anomalous vertices, while the loops themselves are also second or third order and do not allow mixing either; because one would not be able to close them if that was the case. 

\subsection{Direct Scheme}

\begin{figure*}[t!]
\includegraphics[width=0.95\textwidth]{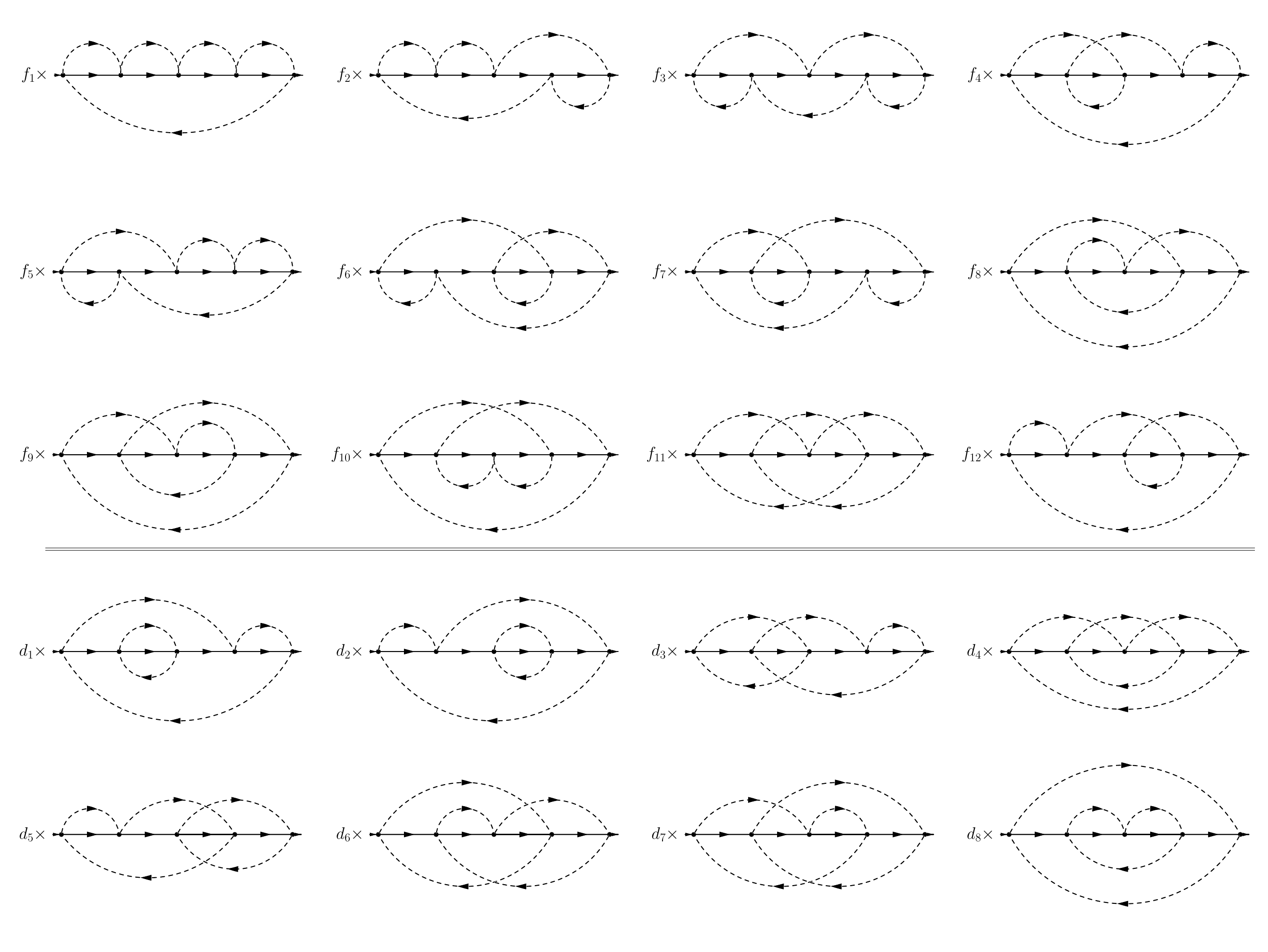}
\caption{Fifth-order diagrams (focusing only on connectivity) contributing to the fermionic self-energy. There are two classes of diagrams, those with a single pseudo\-fermion loop and those with two pseudo\-fermion loops (we separated them by a horizontal line in the figure). The diagrams are the same in all the three schemes, and what changes (from one scheme and vertex assignment to another) are the multiplicative factors, labeled here by $f_{n}$ and $d_{m}$, with $n \in \{1, \dots , 12 \}$, and $m \in \{1, \dots , 8 \}$. Coupling-constant wise, there are only two contributions to the imaginary part of the retarded self-energy (the $J_{\perp}^4 J_{z}$ and $J_{z}^3 J_{\perp}^2$ contributions), and a different set of $f_{n}$'s and $d_{m}$'s is needed for each of the two.}
\label{fig:5thXloopSE1}
\end{figure*}

The relevant diagrams contributing to the fifth-order self-energy are given in Fig.~\ref{fig:5thXloopSE1}, where we show the general form of the diagrams and keep their multiplicative factors as scheme and vertex-label dependent. There will be only two types of vertex-label contributions to the fifth-order self-energy, namely, the $J_{\perp}^4 J_{z}$ and the $J_{z}^3 J_{\perp}^2$ ones. Their multiplicative factors are collected and summarized in Tables~\ref{table:MfactsD} and \ref{table:MfactsC} for the direct and conventional schemes, respectively.
It is important to note that the reason why $f_{11}\!=\!0$ in the direct scheme is not some fortuitous cancellation between different combinations of vertices; it is due to the fact that in that case one is not able to construct such a contraction, since one would need to connect with pseudo\-fermion propagators vertices of the same spin-flip ``polarity''. As in the case of the lower-order calculations, we are showing only the \textit{hole} diagrams with the implicit understanding that we have calculated both \textit{hole} and \textit{particle} contributions, and the calculations proceed in a similar way. Although the integrals in Matsubara time are now somewhat more complicated, the real incremental challenge was in calculating the principal-value energy integrals to obtain the $\tilde{\omega}$ dependence of the imaginary-part of the retarded self-energy. 
(It required lengthier computations as well as cross-checks with numerical integration to make sure that the principal value was evaluated correctly each time.) 
Additionally, before the energy integrals, the $\tau$-integration results were compared to the flat-band limit results for which the exact solution up to arbitrary order in the coupling constant is available via exact diagonalization \cite{Ljepoja2024a}.

\begin{table}[t]
\centering
\begin{tabular}{|c|c|c|c|}
\hline
\multicolumn{4}{|c|}{$J_{\perp}^2 J_{z}^3$ contribution} \\
\hline
 $  \ f_{1}=-10 \ $ & $ \ f_{2}=4 \ $ & $ \ f_{3} =-2\ $ & $ \ f_{4}=-4 \ $  \\
\hline
$ \ f_{5} =4\ $ &$ \ f_{6} =-2\ $ & $ \ f_{7} =-2\ $ & $ \ f_{8}=-2 \ $ \\ 
\hline     
 $ \ f_{9}=2 \ $ & $ \ f_{10}=4 \ $ & $ \ f_{11} = 0 \ $ &$ \ f_{12} =4 \ $  \\
 \hline
$ \ d_{1} =-6\ $ &$ \ d_{2} =6\ $ & $ \ d_{3} =-2\ $ & $ \ d_{4}=-2 \ $ \\ 
\hline     
 $ \ d_{5}=-2 \ $ & $ \ d_{6}=-2 \ $ & $ \ d_7 = -2 \ $ &$ \ d_8 =6 \ $  \\
\hline \hline
\multicolumn{4}{|c|}{$J_{\perp}^4 J_{z}$ contribution} \\
\hline
 $  \ f_{1}=-5 \ $ & $ \ f_{2}=2 \ $ & $ \ f_{3} =-1\ $ & $ \ f_{4}=-2 \ $  \\
\hline
 $ \ f_{5} =2\ $ &$ \ f_{6} =-1\ $& $ \ f_{7} =-1\ $ & $ \ f_{8}=-1 \ $ \\ 
\hline     
 $ \ f_{9}=1 \ $ & $ \ f_{10}=2 \ $ & $ \ f_{11} = 0 \ $ &$ \ f_{12} =2 \ $  \\
\hline
$ \ d_{1} =-3\ $ &$ \ d_{2} =3\ $ & $ \ d_{3} =-1\ $ & $ \ d_{4}=-1 \ $ \\ 
\hline     
 $ \ d_{5}=-1 \ $ & $ \ d_{6}=-1 \ $ & $ \ d_7 = -1 \ $ &$ \ d_8 =3 \ $  \\
 \hline
\end{tabular}
\caption{Multiplicative factors for the direct-scheme calculation of the self-energy. They are given for the only two possible vertex combinations, $J_{\perp}^2 J_{z}^3$ and $J_{\perp}^4 J_{z}$.}
\label{table:MfactsD}
\end{table}

Doing the calculations in the direct scheme, \textit{i.e.} with the factors from Table~\ref{table:MfactsD}, one arrives at the expression for the fifth-order logarithmic contributions to the imaginary-part of the retarded self-energy to be
\begin{equation}
\begin{split}
\Sigma^{i}_{(5,0)a} = & \ C_{1} - \frac{11}{8} \pi^3 J_{\perp}^{4}J_{z} \rho_{0}^{4} \ln(\tilde{\omega}) \\
& \qquad \qquad + 16  \pi J_{\perp}^{4} J_{z} \rho_{0}^{4}  \ln^3(\tilde{\omega})\\
\\
\Sigma^{i}_{(5,0)b} = &  \ C_{2} -\frac{11}{16} \pi^3 J_{\perp}^{2}J^{3}_{z} \rho_{0}^{4} \ln(\tilde{\omega}) \\
& \qquad \qquad + 8 \pi J_{\perp}^{4} J_{z} \rho_{0}^{4}  \ln^3(\tilde{\omega})
\label{eq:5thDcont}
\end{split}
\end{equation}
where we have used labels $C_{1}$ and $C_{2}$ to denote some numerical constants (independent of $\tilde{\omega}$) \footnote{We find: $C_{1} =\frac{8 \pi^4}{3} - 8 i \pi^3 \mathrm{ln}(3) + 2 \pi^2 \big [-2 \ \mathrm{ln}^{2}(3) - 6 \ \mathrm{Li}_{2}(3) + \mathrm{Li}_{2}(9) \big ] + \frac{1}{4} i \pi \big [48 \ \mathrm{ln}(3)  \ \mathrm{Li}_{2}(3) - 4 \ \mathrm{ln}(9) \ \mathrm{Li}_{2}(9) - 48 \ \mathrm{Li}_{3}(3) + 4 \ \mathrm{Li}_{3}(9)+ 75 \ \zeta (3) \big]$, and $C_{2} = \frac{1}{2} \ C_{1}$. The $\mathrm{Li}_{n}(z)$ are poly\-logarithm functions and $\zeta(z)$ is the Riemann zeta function.}, which will play no role in the CS calculation to find the fourth-order terms in the beta functions. Proceeding further with the calculation becomes rather complicated in the generic spin-anisotropic case, as it involves an under\-determined system of equations that requires additional input to solve it (see Appendix \ref{Apdx:anisotropic4th}). On the other hand, the spin-isotropic limit is solvable using only the linear system one obtains from the CS equation alone. 

We shall thus discuss the spin-isotropic calculation here and then come back and remark on the spin-anisotropic case. The series ansatz for the beta function is a bit different than what we had originally for the lower-order calculation. The reason is that we only want to find the channel-number-independent part of the additional term in the fourth-order beta function. In order to do that, we have to explicitly include the channel number into the ansatz,
\begin{equation}
\beta(\mathrm{g}) = a_{1} \mathrm{g}^2 + a_{2} \mathrm{g}^3 + a_{2M} M \mathrm{g}^3 + a_{3} \mathrm{g}^4 
\end{equation}
where we separated explicitly the channel-independent parts from the channel-dependent one. Another way to look at why the ansatz is set up like that, is to stress that we did not calculate the fifth-order diagrams with fermion loops. This means that, to extract from CS equations for the ansatz coefficients that are linearly independent, the highest order we can go to is $M \mathrm{g}^4$. So we need to have a way of extracting coefficients in powers of both $M$ and $\mathrm{g}$, and that is what the proposed ansatz enables us to do. 

As far as the self-energy is concerned, we have already calculated lower-order contributions in the previous sections, and in the spin-isotropic limit, including now also the fifth-order contribution, it reads

\begin{widetext}
\begin{alignat}{5}
\frac{4 \rho_{0}}{\pi}\Sigma^{i} \ = \  P_{(2,0)} \mathrm{g}^{2} & + & \ P_{(3,0)} \  \mathrm{g}^{3} \ \mathrm{\ln} (\tilde{\omega}) & \quad + \ P_{(4,0)} \  \mathrm{g}^{4} \ \mathrm{ln}^{2} (\tilde{\omega}) & \quad + & \ P_{(5,0a)}  \  \mathrm{g}^{5} \ \mathrm{\ln}^{3} (\tilde{\omega}) && &  \nonumber \\
 & & & \quad + \ P_{(4,1b)} \ M \ \mathrm{g}^{4} \ \ln(\tilde{\omega}) & \quad + & \ P_{(5,1a)}  \ M \ \mathrm{g}^{5} \ \mathrm{\ln}^{2} (\tilde{\omega})&& & \nonumber \\
& & & \quad +  \ P_{(4,1a)} \ M \ \mathrm{g}^{4} & \quad + & \ \big [ P_{(5,0b)} + P_{(5,1b)} \ M \big ]  \  \mathrm{g}^{5} \ \mathrm{\ln} (\tilde{\omega}) &&&\nonumber \\
&&& & \quad +& \ P_{(5,1c)} \ M \ \mathrm{g}^{5} \ &&& \nonumber \\
&&&&&&&&
\label{eq:SE5thd}
\end{alignat}
\end{widetext}
We have introduced labels here also for the contributions from the fifth-order diagrams with fermion loops (the channel-number-dependent ones), but in our present calculation we will disregard them. Interested readers are referred to Appendix \ref{Apdx:FullBetaIsotropic4th} to see the calculation redone in the spin-isotropic limit with the contribution of the fermion-loop diagrams included. The needed coefficients of the self-energy are given by
\begin{equation}
    \begin{split}
    &P_{(2,0)} = 3 \\
    &P_{(3,0)} = -12 \\
    &P_{(4,0)} = 36 \\
    &P_{(4,1a)} = 24 [\mathrm{ln}(2) - 1] \\
    &P_{(4,1b)} =24 \\
    &P_{(5,0a)} = -96 \\
    &P_{(5,0b)} = 12 \pi^2
    \end{split}
\end{equation}

One proceeds now in the same way as we did in the case of the lower-order calculation. Namely, we replace into the CS equation and extract from it equations that connect the $P$ coefficients of the self-energy with the coefficients of the beta-function ansatz. The set of equation one arrives at are
\begin{equation}
    \begin{split}
        a_{1} &= \frac{P_{(3,0)}}{2 P_{(2,0)}} \ ,
        \qquad a_{2} = 0\\
        a_{2M} &= \frac{P_{(4,1b)}}{2 P_{(2,0)}} \ ,
        \qquad a_{3} = \frac{P_{(5,0b)}}{2 P_{(2,0)}}
    \end{split}  \label{eq:eqD5th}  
\end{equation}
as well as a set of RG-consistency equations
\begin{equation}
    a_{1} = \frac{2 P_{(4,0)}}{3 P_{(3,0)}} \ , \qquad a_{2} = 0 \ , \qquad a_{1} = \frac{3 P_{(5,0a)}}{4 P_{(4,0)}}
    \label{eq:eqD5thc}
\end{equation}
which are automatically satisfied by the solutions of Eqs.~(\ref{eq:eqD5th}) with the given values of the $P$ coefficients. 

This way we are able to calculate the beta function (with a channel-number-independent fourth-order term) is
\begin{equation}
\begin{split}
\beta &= -2 \mathrm{g}^2 + 4 M \mathrm{g}^3 + 2 \pi^2 \mathrm{g}^4\\
&\mapsto -g^2+ M g^3 +\frac{\pi^2}{4}g^4
\end{split}
\end{equation}
This is the RG flow in the  spin-isotropic limit of the (multi) two-channel Kondo model (taken beyond the standard first two terms). It was a straightforward calculation once the self-energy is given, since it is unequivocally determined by the CS equation.

\subsection{Conventional scheme}

\begin{table}[b]
\centering
\begin{tabular}{|c|c|c|c|}
\hline
\multicolumn{4}{|c|}{$J_{\perp}^2 J_{z}^3$ contribution} \\
\hline
 $  \ f_{1}=-8 \ $ & $ \ f_{2}=0 \ $ & $ \ f_{3} =-8\ $ &$ \ f_{4}=-4 \ $  \\
\hline
 $ \ f_{5} =0\ $ &$ \ f_{6} =-4\ $ & $ \ f_{7} =-4\ $ & $ \ f_{8}=-4 \ $ \\ 
\hline     
 $ \ f_{9}=4 \ $ & $ \ f_{10}=0 \ $ & $ \ f_{11} = -8 \ $ &$ \ f_{12} =4 \ $  \\
 \hline
$ \ d_{1} =-4\ $ &$ \ d_{2} =4\ $ & $ \ d_{3} =4\ $ & $ \ d_{4}=0 \ $ \\ 
\hline     
 $ \ d_{5}=4 \ $ & $ \ d_{6}=4 \ $ & $ \ d_7 = 4 \ $ &$ \ d_8 =8 \ $  \\
\hline \hline
\multicolumn{4}{|c|}{$J_{\perp}^4 J_{z}$ contribution} \\
\hline
 $  \ f_{1}=-12 \ $ & $ \ f_{2}=4 \ $ & $ \ f_{3} =-4\ $&$ \ f_{4}=0 \ $   \\
\hline
$ \ f_{5} =4\ $ &$ \ f_{6} =-8\ $ & $ \ f_{7} =-8\ $ & $ \ f_{8}=-8 \ $\\ 
\hline     
 $ \ f_{9}=8 \ $ & $ \ f_{10}=4 \ $ & $ \ f_{11} = -12 \ $ &$ \ f_{12} =0 \ $  \\
 \hline
$ \ d_{1} =-12\ $ &$ \ d_{2} =12\ $ & $ \ d_{3} =4\ $ & $ \ d_{4}=8 \ $ \\ 
\hline     
 $ \ d_{5}=4 \ $ & $ \ d_{6}=4 \ $ & $ \ d_7 = 4 \ $ &$ \ d_8 =8 \ $  \\
\hline   
\end{tabular}
\caption{Multiplicative factors for the self-energy diagrams in the conventional-scheme calculation.}
\label{table:MfactsC}
\end{table}

In the conventional scheme, the calculation will proceed in the same way as in the direct one. However, in this scheme and as was the case for the lower-order calculation, we are going to have unphysical contributions to the self-energy. Moreover, at this order there will be a larger number of diagrams that do not have a valid translation in terms of the direct-scheme fermions and processes. All these unphysical contributions come from the $J_{\perp}^4 J_{z}$ diagrams. This is as expected, considering that in those diagrams there are more than two spin-flip vertices present, which allows for the mixing of normal and anomalous ones.

\begin{figure}[t]
    \centering
    \includegraphics[width=0.48\textwidth]{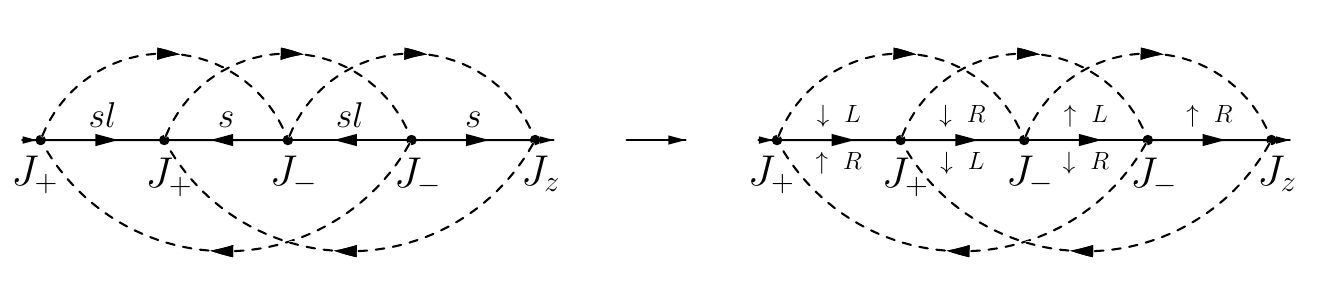}
    \caption{Example of an unphysical contribution to the electron self-energy in the conventional scheme and its attempted translation to the direct scheme. The double label assignments on the right diagram highlight the need for (unphysical) off-diagonal propagators to represent those processes in the original direct language.}
    \label{fig:untranslatable5thD12}
\end{figure}

As already mentioned in the previous section, the set of available diagrams is going to be the same in all schemes (shown in Fig.~\ref{fig:5thXloopSE1}) but with different multiplicative factors. For the conventional scheme, the multiplicative factors are given in Table~\ref{table:MfactsC}. There are two main differences when comparing the \textit{conventional} multiplicative factors with the \textit{direct} ones (given in Table~\ref{table:MfactsD}). 
The first is that the $f_{11}$ coefficient is no longer zero. For the $J_{\perp}^2 J_{z}^3$ contributions, this is a direct consequence of the fact that the $J_{z}$ vertex scatters only $s$-sector fermions. On the other hand, for the $J_{\perp}^4 J_{z}$ contribution, this diagram appears because in the conventional scheme one is able to connect two spin-flip vertices of the same polarity  (either $J_{+}$ with $J_{+}$ or $J_{-}$ with $J_{-}$) with a single fermion propagator, which is a consequence of allowing for the mixing of normal and anomalous vertices (see Fig.~\ref{fig:untranslatable5thD12} for an example). 
For a systematic comparison, we can distinguish the unphysical contributions as ``fully'' and ``partially'' unphysical. Partially unphysical diagrams are untranslatable only when they mix normal and anomalous vertices, while in all other cases they have a valid translation in terms of the original (direct-scheme) fermions. Fully unphysical diagrams, in contrast, are the ones that are untranslatable for any combination of vertices. These are diagrams that cannot be made at all in the direct scheme, but they exist in the conventional scheme and always mix normal and anomalous vertices. 
At this order, we encounter the first appearance of a fully unphysical diagram (the one we selected for Fig.~\ref{fig:untranslatable5thD12}). 
We can see that the valid conventional-scheme diagram would translate into one that violates conservation of both ``lead'' and ``spin'' along the propagators in the direct-scheme diagram. 
The second noticeable difference between Tables \ref{table:MfactsD} and \ref{table:MfactsC} is that, in the conventional scheme, a number of other multiplicative factors are zero. In fact, there are no other zeros in the direct scheme besides $f_{11}$. Once again, the reason for the appearance of these zeros in the conventional scheme is the nonexistence of $sl$-fermion scattering from the $J_{z}$ vertices. 

The calculation now proceeds in an identical way as in the direct scheme. The same as in the lower-order case, we arrive at an expression for the imaginary part of the retarded self-energy of the same form as that in Eq.~(\ref{eq:SE5thd}) but with different coefficients given by
\begin{equation}
    \begin{split}
    &P_{(2,0)} = 8 \\
    &P_{(3,0)} = -32 \\
    &P_{(4,0)} = 96 \\
    &P_{(4,1a)} = 64 [\mathrm{ln}(2) - 1] \\
    &P_{(4,1b)} =64 \\
    &P_{(5,0a)} = -256 \\
    &P_{(5,0)b} = \frac{68}{3} \pi^2
    \end{split}
\end{equation}
The beta-function ansatz is still the same, so the equations one finds for their coefficients are the same as well. The RG-consistency equations are still satisfied; however, we arrive at a different channel-independent fourth-order term in the RG-flow than we did in the direct scheme. Namely, the conventional beta function is found to be
\begin{equation}
\begin{split}
\beta &= -2 \mathrm{g}^2 + 4 M \mathrm{g}^3 +\frac{11  \pi^2}{8} \mathrm{g}^4\\
&\mapsto -g^2+ M g^3 +\frac{11 \pi^2}{64}g^4
\end{split}
\end{equation}
Comparing results, the conventional scheme gives a fourth-order coefficient that is about one-and-a-half (precisely $1.\overline{45}\ldots\!=\!16/11$) times smaller than the one from the direct-scheme, --a consequence of the presence of unphysical processes among the conventional diagrams. 

\subsection{Consistent scheme}

The consistent scheme produces the same result as the direct one. In this scheme, unphysical diagrams are precluded, since the mixing of anomalous and normal vertices is inhibited by the $\tilde{n}$ factors. The result for the self-energy is the same as given in Eq.~(\ref{eq:consSE}). In fact, the results coincide with the direct-scheme ones to all orders in the coupling constant (and, as previously discussed, it can be obtained using either the $s$- or the $sl$-fermion self-energy).

\subsection{Comparison of results beyond third order}

\begin{figure}[b!]
    \centering
    \includegraphics[width=0.23\textwidth]{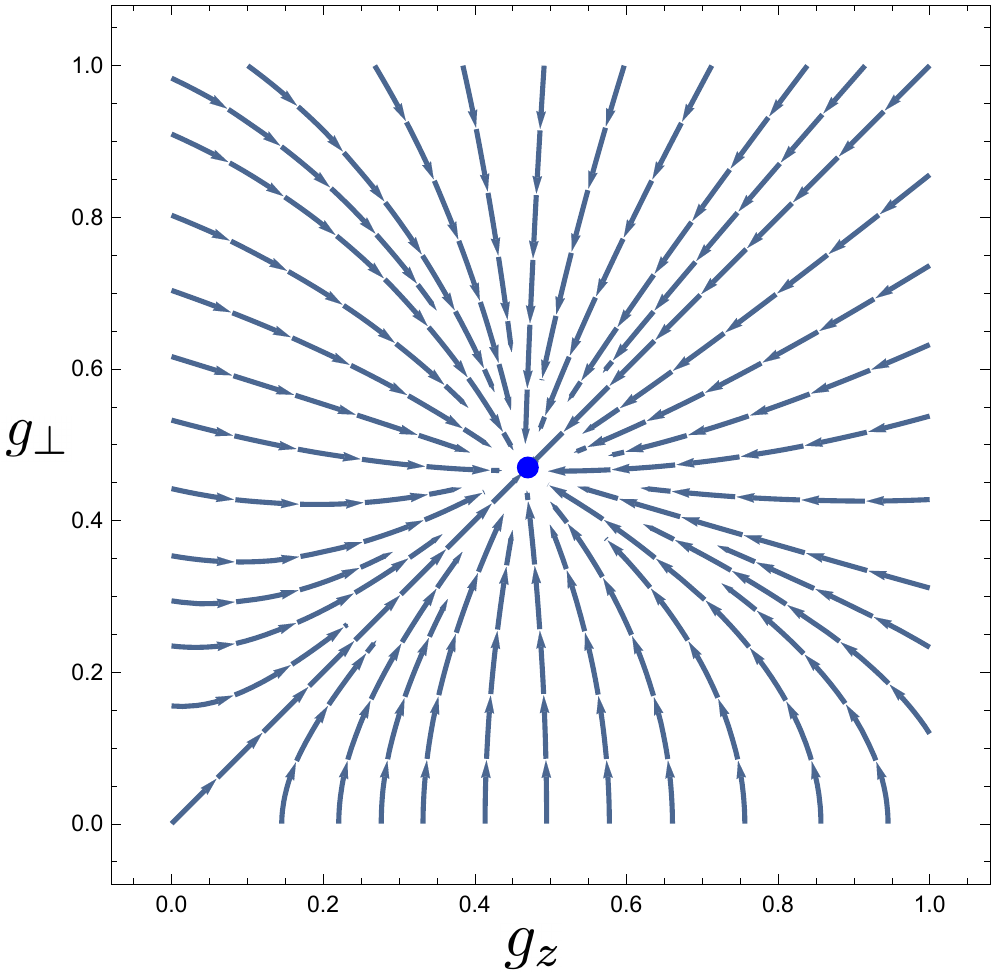}
    \includegraphics[width=0.2295\textwidth]{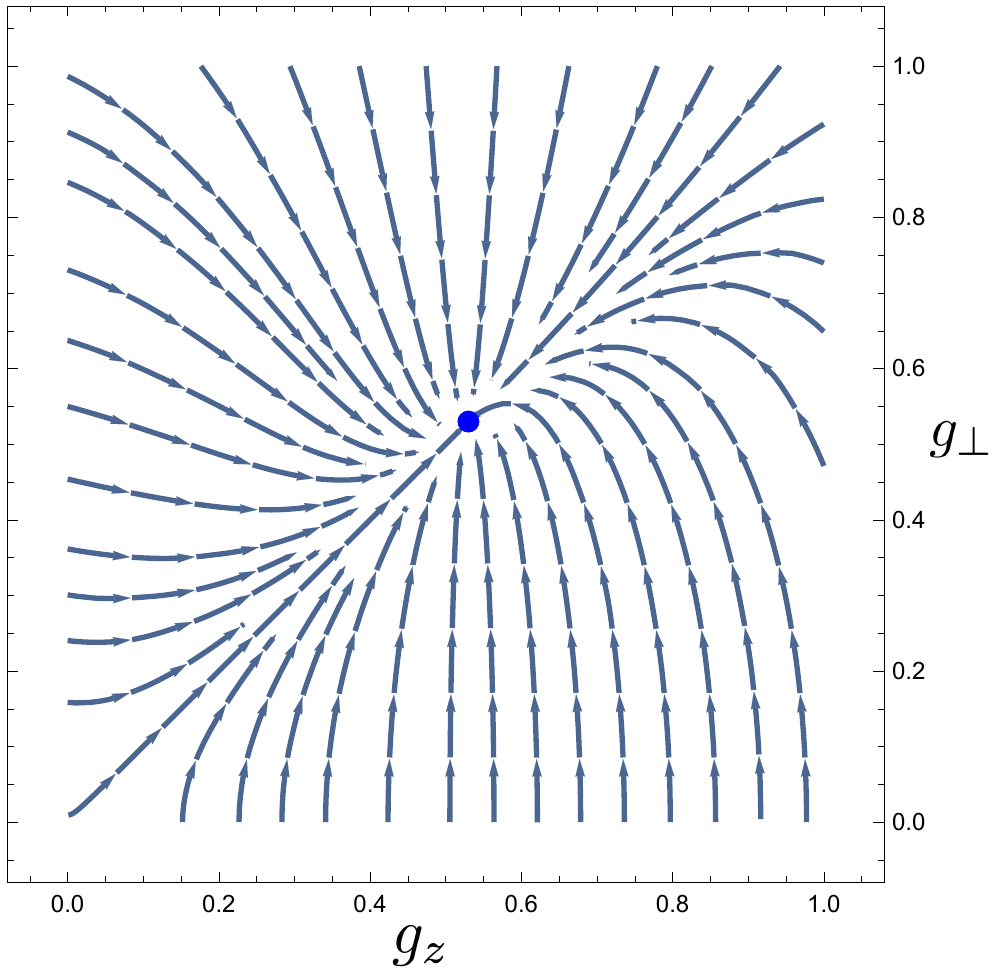}
    \caption{Comparison of the RG flow in the direct (left panel) and conventional (right panel) schemes for $M=1$. In the \textit{beyond-third-order} flow, the position of the fixed point in the direct scheme shifts to $g_{\perp} = g_{z} \simeq 0.47$, and for the conventional scheme to $g_{\perp} = g_{z} \simeq 0.53$, (but be reminded that this flow diagrams exaggerate the differences, because they include only $M$-independent contributions at the fourth order; including all --see Appendix~\ref{Apdx:FullBetaIsotropic4th}-- the fixed points would shift further, to $0.38$ and $0.42$, respectively).}
    \label{fig:flowcomparison}
\end{figure}

In conclusion, we have shown that the presence of what we dubbed ``unphysical diagrams'' in the conventional scheme causes a discrepancy between the RG flows calculated in the conventional scheme as opposed to the direct (or consistent) one. This discrepancy between schemes is more obvious in the spin-an\-isotropic case (results and details of that calculation are given in Appendix~\ref{Apdx:anisotropic4th}). In Fig.~\ref{fig:flowcomparison}, we show the graphic comparison of the flows in the two schemes. One can see that, in the conventional scheme, the fixed point occurs at a higher value of the coupling than in the direct scheme. Moreover, the flow itself is similar in the two schemes above and along the diagonal (which represents the spin-isotropic flow), but below the diagonal they differ significantly in the way they approach the respective fixed points. However, this discrepancy is caused by the channel-number-independent terms in the beta function, which in the large-channel-number limit would be disregarded. Below, we shall investigate what happens with the comparison of the different schemes in such a limit.

\subsubsection{Large-$M$ limit}

\begin{figure}[t]
    \centering
    \includegraphics[width=0.48\textwidth]{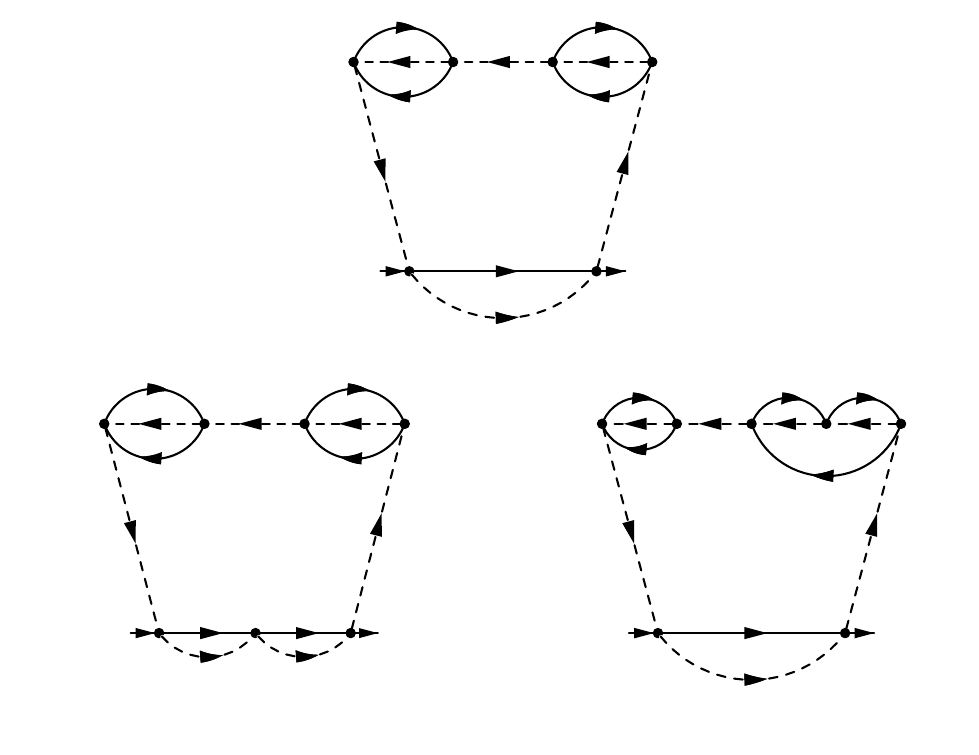}
    \caption{Examples of sixth- and seventh-order diagrams that contribute to the self-energy in the large-$M$ limit. The diagram in the first row is a sixth-order diagram with two fermionic loops [which could be labeled as $(6,2)$] and thus gives an $M^2$ contribution to the self-energy. The two bottom diagrams are both seventh-order ones with two fermionic loops [labeled as $(7,2)$], and they also give $M^{2}$ contributions. All of these diagrams produce always ``physical'' contributions (\textit{i.e.}, ``translatable'' or not unphysical) when in the conventional scheme.}
    \label{fig:largeM}
\end{figure}

Calculating the beta function in the large-channel-number limit would proceed by keeping, at each order in the coupling-constant ($\mathrm{g}$) expansion of the self-energy, only the highest powers of the channel number ($K\!=\!2M$) that are possible at that order (and in accordance with the order of the calculation). In other words, one keeps the self-energy diagrams that have the largest number of fermion loops at the last kept order in the coupling constant expansion. Figure~\ref{fig:5thloopSE} shows examples of such diagrams for the fifth order; they are effectively fourth-order contributions in the large-$M$ approximation (since, near the fixed point, each power of $M$ would lower the order of the diagram by one \cite{gan1994}). Other fifth-order diagrams with no fermion loops are then disregarded (\textit{i.e.}, those in Fig.~\ref{fig:5thXloopSE1} that were responsible for the discrepancies discussed above).
In such a calculation, one would have to proceed also to the sixth order as well, but keep only the diagrams that have two fermion loops (disregarding the ones with no or one fermion loop only), since they also represent effective fourth-order contributions; --see Fig.~\ref{fig:largeM} for an example of such diagrams. 

The same logic applies to the seventh-order contributions, one would only keep the diagrams with two fermion loops (\textit{i.e.}, effective fifth-order diagrams). Examples of those higher-order contributions are also shown in Fig.~\ref{fig:largeM}, --where we refrain from showing all of the other possible pseudo\-fermion contractions to keep the diagrams simple, since all we are interested in are the fermion propagators and their loops. At this point the question arises whether or not to include other fifth-order diagrams, such as those from Fig.~\ref{fig:5thXloopSE1}, that would reintroduce the discrepancies among schemes. The choice to exclude them is justified if one is interested in the (``$M$ainstem'') RG-flow from weak coupling and constitutes an additional prescription for how the large-$M$ limit is defined
\footnote{Notice that with this definition one would skip also the no-loop fourth-order diagrams, but those did not contribute to the beta functions anyway. They contributed to the RG-consistency conditions, which get nevertheless reorganized in the large-$M$ limit.}; cf.~Ref.~\onlinecite{fischer1999} and see also Ref.~\onlinecite{DEla1990} and Sec.~\ref{Sec:extRG}.

By looking at the diagrams in Figs.~\ref{fig:5thloopSE} and \ref{fig:largeM} we see that, for the purpose of translation among schemes, these diagrams can be separated into two parts. One is the set of loops themselves. Such loops will always be of second or third order in the coupling. One can make higher-order loops for higher-order diagrams, but such diagrams shall be neglected since, the way we defined the large-$M$ limit, at each order we keep only the diagrams with the largest possible number of loops.
The second- and third-order loops do not mix the anomalous and normal vertices. A second-order loop cannot, because if it did one would not be able to make a loop (one would have to contract two fermionic creation/annihilation operators in order to close a mixed loop). A third-order loop always involves the $J_{z}$ vertex and only two $J_{\perp}$ ones, so the same argument as for the second-order loops still holds. 
The other part of the diagram (its ``baseline'') is effectively like either a second- or third-order self-energy diagram (with dressed pseudo\-fermions), since one had to reserve as many vertices as possible to make as many loops as possible. In such second- and third-order diagrams there is no mixing of normal and anomalous vertices if one seeks to calculate the $s$-sector self-energy. Trivial mixing, where a loop might be with all anomalous vertices and the baseline or another loop with all normal ones, or vice versa, does not produce any unphysical (untranslatable) contributions. As a result, this analysis reaches the conclusion that all of the diagrams in the large-$M$ limit (up to arbitrarily high order in the coupling) are going to be translatable from the conventional to the direct scheme. Therefore, we do not expect any discrepancies between the beta functions obtained from two schemes in this limit. 

This and the previous analyses point to the limit of applicability of the scheme based on the conventional compactification of the model. In an appropriately defined large-$M$ limit, it always produces a ``physical'' result to all orders of the perturbative expansion. But away from the large-$M$ approximation, it is only ``physical'' for low orders in the expansion (going beyond a third-order expansion in the coupling constant, the beta function already includes the effects of unphysical contributions).

\section{Extended RG Analysis and Universality}
\label{Sec:extRG}

So far we have seen that a difference arises between the direct (or consistent) and conventional schemes when going beyond the third order in the flow equations of the (multi) two-channel Kondo model. It can be argued, however, that at this order the beta functions are already nonuniversal (and thus, from the point of view of the universal aspects of the physics, all three schemes are interchangeable). Indeed, in the ``conventional lore'' (invoking the scaling limit and comparing theories differing simply on the cutoff scale at which the bare coupling is defined), the universal parts of the beta function are only its $g^2$ and $g^3$ terms, for which all ``schemes'' give the same result. This assumes that the couplings in the two schemes are related in a flow-restricted way (see below). More general schemes, some going beyond the perturbative definition of the theories, could have their couplings related in more general (even non\-analytic) ways \cite{andrei1984} and then the universality could be even further restricted to the $g^2$ term only
\footnote{The situation is different for the (multi\-channel) single-impurity Anderson model (SIAM) counterparts, for which the impurity thermodynamics is finite and can be calculated without the need to introduce a cutoff scale for the bulk \cite{bolech2002,*bolech2005a}. Those models shall be discussed elsewhere.}. 
Our comparison of the \textit{direct} and \textit{indirect} schemes, connected to the direct one by BdB-based compactification procedures, goes beyond the simple situation of two theories related purely by scaling. However, in all these cases, the theories are perturbatively defined and share a similar UV Gaussian fixed point with a decoupled impurity, so the correspondence among beta functions extends to the $g^3$ terms as well, as we indeed found. Remarkably, the agreement between the direct- and consistent-scheme beta functions extends even beyond, and to all orders of perturbation theory, underlining the fact that the \textit{consistent} BdB procedure is an exact mapping not circumscribed to the scaling limit.

Let us discuss in more detail the universal aspects of scaling in perturbatively defined theories as applied to the results of the previous sections. We start from the spin-an\-isotropic beta functions given by these arbitrary-coefficient expressions
\begin{equation}
\begin{split}
& \beta_{\perp}(g_{z}, g_{\perp}) =  a_{1} g_{z}g_{\perp}+ a_{2}  g_{\perp}^3+ a_{3}g_{\perp} g_{z}^2\\
&  \beta_{z}(g_{z}, g_{\perp}) = a_{1} g_{\perp}^2+  (a_{2} + a_{3}) g_{\perp}^2 g_{z}
\end{split}
\end{equation}
(notice they coincide in the spin-isotropic limit, $g_{z}\!=\!g_{\perp}\!=\!g$, in which we recover the single-coupling-constant case extensively discussed in the literature). Performing a change of the variables of the form: $g_{\perp} = \bar{g}_{\perp} + c_{2} \bar{g}_{\perp} \bar{g}_{z}$ and $g_{z} = \bar{g}_{z} + c_{2} \bar{g}^{2}_{\perp}$, ---where (i) the two sets of coupling constants coincide at the ``tree level'' since there are no divergences and thus no ambiguities in their definitions, and (ii) the next terms in the mapping are dictated by the leading order of the respective RG flows \cite{Stevenson1981,andrei1984,Aristov2009} (in accordance with the \textit{large-river} picture that focuses on the main\-stem flow and ignores ``(dis)tributaries'' associated with microscopic --and other, like large-$M$ deviation-- details \cite{Delamotte2012,*Delamotte2004})---, we arrive at the same beta functions for the redefined couplings
\begin{equation}
\begin{split}
& \bar{\beta}_{\perp}(\bar{g}_{z}, \bar{g}_{\perp}) \equiv \frac{d \bar{g}_{\perp}}{d \mathrm{ln}(D)}  = a_{1} \bar{g}_{z}\bar{g}_{\perp}+ a_{2}  \bar{g}_{\perp}^3+ a_{3}\bar{g}_{\perp} \bar{g}_{z}^2\\
&  \bar{\beta}_{z}(\bar{g}_{z}, \bar{g}_{\perp}) \equiv \frac{d \bar{g}_{\perp}}{d \mathrm{ln}(D)} = a_{1} \bar{g}_{\perp}^2+  (a_{2} + a_{3}) \bar{g}_{\perp}^2 \bar{g}_{z}
\end{split}
\end{equation}
But notice that this is true only for the beta function(s) up to third order. Going beyond the third order, the RG flow is no longer universal, in the sense that a change of variables will lead to different fourth-order coefficients that will depend on the arbitrary parameter in the change of variables ($c_2$). Or, differently put, an ``error'' in identifying the position along the flow at which the bare coupling constants are defined affects the beyond-third-order details of that flow.

We now ask the reader to recognize that this universality of the third-order beta function would be retained even if the beta function were to be $\tilde{\omega}$ dependent (via its coefficients $a_{1,2,3}$). This is important, because next we want to compare the $\tilde{\omega}$ (or cutoff) dependent beta functions obtained by direct- and conventional-scheme calculations up to third order in an extended-RG sense. They will be obtained using the same CS method as earlier, but instead of keeping only constant and divergent [power-of-$\log(\tilde{\omega})$] terms, one allows also for integer powers of $\tilde{\omega}$ in the self-energy. Physically, this corresponds to stretching beyond the IR-fixed-point-dominated region and into the low-$T$ part of the crossover regime of the Kondo model. We argue that certain aspects of that regime are also universal in the same sense as above.

We thus recalculate the self-energy, but this time keeping and reorganizing the expression in powers of $\tilde{\omega}$. We have $\tilde{\omega}^{0}$ (those terms already calculated), as well as $\tilde{\omega}^{2}$ and $\tilde{\omega}^{4}$ contributions, etc. There are no odd powers of $\tilde{\omega}$ in the series expansion of the self-energy. The diagrams and calculations are the same as they were before, it is just that in doing the series expansion of the final contribution for each of them we go beyond the dominant contribution and keep also the first few subdominant terms. We can write a formal expansion for the self-energy as
\begin{equation}
\Sigma^{i}(\tilde{\omega}) = \Sigma^{i}_{0}(\ln(\tilde{\omega})) + \tilde{\omega}^2 \Sigma^{i}_{2}(\ln(\tilde{\omega})) + \tilde{\omega}^4 \Sigma^{i}_{4}(\ln(\tilde{\omega})) + \ldots
\label{eq:subSE}
\end{equation}
where $\Sigma^{i}_{0}$ is the $\ln(\tilde{\omega})$-dependent contribution given by the expression in Eq.~(\ref{eq:SEd}), while $\Sigma^{i}_{2}$ and $\Sigma^{i}_{4}$ are the alike $\tilde{\omega}^2$ and $\tilde{\omega}^4$ coefficients of an integer-frequency-powers reorganization of the expansion, respectively. In the direct scheme they are given by
\begin{equation}
\begin{split}
    & \frac{4 \rho_{0}}{\pi} \Sigma^{i}_{2} = P^{\perp z}_{(3,0)(2)} \ \mathrm{g}_{\perp}^2 \mathrm{g}_{z} + P^{\perp }_{(4,0)(2)} \ \mathrm{\ln}(\tilde{\omega}) \ \mathrm{g}_{\perp}^4 \\
    & \qquad \qquad \qquad + P^{\perp z}_{(4,0)(2)} \ \mathrm{\ln} (\tilde{\omega}) \ \mathrm{g}_{\perp}^2 \mathrm{g}_{z}^2\\
    & \frac{4 \rho_{0}}{\pi} \Sigma^{i}_{4} = P^{\perp z}_{(3,0)(4)} \ \mathrm{g}_{\perp}^2 \mathrm{g}_{z} +  \big [ P^{\perp }_{(4,0)(4a)} + P^{\perp }_{(4,0)(4b)} \ \mathrm{\ln}(\tilde{\omega}) \big ] \ \mathrm{g}_{\perp}^4 \\
    & \qquad \qquad \qquad +  \ \big [ P^{\perp z}_{(4,0)(4a)} + P^{\perp z}_{(4,0)(4b)} \ \mathrm{\ln} (\tilde{\omega}) \big ] \ \mathrm{g}_{\perp}^2 \mathrm{g}_{z}^2\\
\end{split}
\end{equation}
As one can see, in addition to the $P$ coefficients we had for the logarithmic\-ally divergent part of the self-energy [see Eqs.~(\ref{eq:P3rdDirect}) and (\ref{eq:P3rdConventional}) for the direct and conventional schemes, respectively], we have additional coefficients which we labeled with the powers of $\tilde{\omega}$ they are associated with. For the direct scheme they are
\begin{equation}
\begin{split}
    &P^{\perp z}_{(3,0)(2)} = -6, \qquad P^{\perp }_{(4,0)(2)} = 12, \qquad  P^{\perp z}_{(4,0)(2)}=24 \\
    &P^{\perp z}_{(3,0)(4)} = -3, \qquad P^{\perp }_{(4,0)(4a)} = 3, \qquad P^{\perp }_{(4,0)(4b)} = 6 \\
    &P^{\perp z}_{(4,0)(2a)}=6, \qquad  P^{\perp z}_{(4,0)(2b)}=12
\end{split}
\end{equation}

On the other hand, for the conventional scheme, the self-energy contributions have a slightly different form than in the direct case. One finds,
\small
\begin{equation}
\begin{split}
    & \frac{4 \rho_{0}}{\pi} \Sigma^{i}_{2} = P^{\perp z}_{(3,0)(2)} \ \mathrm{g}_{\perp}^2 \mathrm{g}_{z} +  \big [P^{\perp }_{(4,0)(2a)} + P^{\perp }_{(4,0)(2b)}  \ \mathrm{\ln}(\tilde{\omega}) \big ] \ \mathrm{g}_{\perp}^4 \\
    & \qquad \qquad \qquad +  \big [ P^{\perp z}_{(4,0)(2a)}  + P^{\perp z}_{(4,0)(2b)}  \ \mathrm{\ln} (\tilde{\omega}) \big ] \ \mathrm{g}_{\perp}^2 \mathrm{g}_{z}^2\\
    & \frac{4 \rho_{0}}{\pi} \Sigma^{i}_{4} = P^{\perp z}_{(3,0)(4)} \ \mathrm{g}_{\perp}^2 \mathrm{g}_{z} + \ \big [ P^{\perp }_{(4,0)(4a)}  + P^{\perp }_{(4,0)(4b)} \ \mathrm{\ln}(\tilde{\omega}) \big ] \ \mathrm{g}_{\perp}^4 \\
    & \qquad \qquad  + P^{\perp z}_{(4,0)(4)} \ \big [ P^{\perp z}_{(4,0)(4a)}+ P^{\perp z}_{(4,0)(4b)} \ \mathrm{\ln} (\tilde{\omega}) \big ] \ \mathrm{g}_{\perp}^2 \mathrm{g}_{z}^2\\
\end{split}
\end{equation}
\normalsize
and the coefficients are now given by
\begin{equation}
\begin{split}
    &P^{\perp z}_{(3,0)(2)} = -4, \qquad P^{\perp }_{(4,0)(2a)} = 4, \qquad  P^{\perp z}_{(4,0)(2b)}=8\\
    &P^{\perp z}_{(4,0)(2a)} = -4, \qquad P^{\perp z}_{(4,0)(2b)} = 16, \qquad  P^{\perp z}_{(3,0)(4)}=-2\\
    &P^{\perp}_{(4,0)(4a)} = \frac{14}{3}, \qquad P^{\perp }_{(4,0)(4b)} = 4, \qquad  P^{\perp z}_{(4,0)(4a)}=\frac{4}{3}\\
    &P^{\perp z}_{(4,0)(4b)} = 8
\end{split}
\end{equation}
We can see that the main difference between the conventional- and the direct-scheme subdominant contributions to the self-energy (besides the value of the $P$ coefficients) is in the actual log-power structure of the contributions. In the conventional scheme we have an $\tilde{\omega}^2\ln(\tilde{\omega})$ contribution that is absent in the direct one. 

We can now use the same beta-function ansatz as before [see Eq.~(\ref{eq:beta0})] but for this calculation the coefficients in front of the powers of the coupling constant are now integer-power series in $\tilde{\omega}$,
\begin{equation}
    \begin{split}
       & a_{n} (\tilde{\omega}) = a_{n,0} + a_{n,2} \ \tilde{\omega}^2  + a_{n,4} \ \tilde{\omega}^4 \\
    &  b_{m} (\tilde{\omega}) = b_{m,0} + b_{m,2}  \ \tilde{\omega}^2  + b_{m,4} \ \tilde{\omega}^4
    \end{split}
\end{equation}
where $n=1,2,3$ and $m=1,2,3,4$. Applying the CS equation on the self-energy in Eq.~(\ref{eq:subSE}), we can calculate the above coefficients in the expansion. We treat all the powers of $\mathrm{\ln}(\tilde{\omega})$ and $\tilde{\omega}$ as independent in the CS equation in order to obtain the equations for the coefficients. Carrying out the calculation we arrive at the following \textit{extended beta functions} (including the usually neglected regular cutoff dependence) for the direct scheme:
\begin{equation}
\begin{split}
&\beta_{\perp}= -\big (1 + \tilde{\omega}^2 + \tilde{\omega}^4 \big ) \ g_{\perp}g_{z} +  \frac{M}{2} g_{\perp}^{3} + \frac{M}{2} g_{\perp} g_{z}^2 \\
& \beta_{z} = - \big (1 + \tilde{\omega}^2 + \tilde{\omega}^4 \big ) \ g_{\perp}^2 + M g_{\perp}^2 g_{z}
\end{split}
\end{equation}
We find that, in this scheme, only the lowest-order term in each beta function has any $\tilde{\omega}$ dependence. The rest of the terms stay the same as when we kept only the dominant contributions in the self-energy. 

The result is significantly different in the conventional scheme (with respect to the flow of the spin-flip coupling constant). The beta functions are
\begin{equation}
\begin{split}
&\beta_{\perp}= -\big (1 + \tilde{\omega}^2 + \tilde{\omega}^4 \big ) \ g_{\perp}g_{z} +  \frac{M}{2} \big (1 + 2 \ \tilde{\omega}^2 +\frac{8}{3} \ \tilde{\omega}^4 \big ) \ g_{\perp}^{3} \\
& \qquad \qquad \qquad \qquad + \frac{M}{2} \big (1 - 2 \ \tilde{\omega}^2 -\frac{8}{3} \ \tilde{\omega}^4 \big ) \ g_{\perp} g_{z}^2 \\
& \beta_{z} = - \big (1 + \tilde{\omega}^2 + \tilde{\omega}^4 \big ) \ g_{\perp}^2 + M g_{\perp}^2 g_{z}
\end{split}
\end{equation}

The extended RG flow of $g_{z}$ is the same for both schemes while that of $g_{\perp}$ is different (they are thus different only in the spin-an\-isotropic case). This echoes the fact that the unphysical diagrams are present in only the $g_{\perp}^4$ conventional contributions to the self-energy. This is an interesting result, since the difference between the beta functions obtained via the two different calculation schemes extends now to the universal terms. This helps to further set the limits of applicability of the conventional BdB scheme. At vanishing temperature and for large cutoff (or small frequency) one has that $\tilde{\omega} \rightarrow 0$ and the universal parts of the beta functions are the same as in the direct calculation. On the other hand, if one wants to move away from the low-temperature limit and explore the crossover regime of the model, the conventional scheme starts to be affected by the inclusion of unphysical processes affecting the universal part of the flow (a brief illustration of this is given in Appendix~\ref{Apdx:subdominant}).

\section{Consistent refermionization of the parallel Kondo interaction revisited}

One of the main differences between the consistent and conventional schemes of compactification of the (multi) two-channel Kondo model is that the $J_{z}$ vertex in the conventional scheme only scatters $s$-sector fermions. This results in an asymmetry between the two physical sectors that couple to the impurity in the conventional scheme. As we have seen in the previous sections, this fact forces one to choose the $s$-fermion self-energy to use in the CS equation to determine the beta functions, --as opposed to the consistent scheme in which we are free to choose either-sector self-energy.

We derived the more symmetric form of the compactified parallel Kondo coupling via the use of consistent BdB arguments \cite{Ljepoja2024a}, but they were subtler than in the case of the perpendicular coupling; cf.~Ref.~\onlinecite{shah2016,*bolech2016}. An alternative way to understand the symmetric-footing appearance of $sl$-sector fermions in the $J_{z}$ vertex in the consistent scheme is by noticing that (even if one did not include it in the starting compactified Hamiltonian), already from second-order re\-scaling, parallel scattering of $sl$-sector fermions is generated by the RG flow (provided the $\tilde{n}$ factors in the perpendicular scattering are treated \textit{consistently}). In other words, one can start with the consistently refermionized spin-flip part of the Hamiltonian while leaving the no spin-flip part as in the conventional case, and the model will still flow towards the consistently refermionized model. 

\begin{figure}[t]
    \centering
    \includegraphics[width=0.48\textwidth]{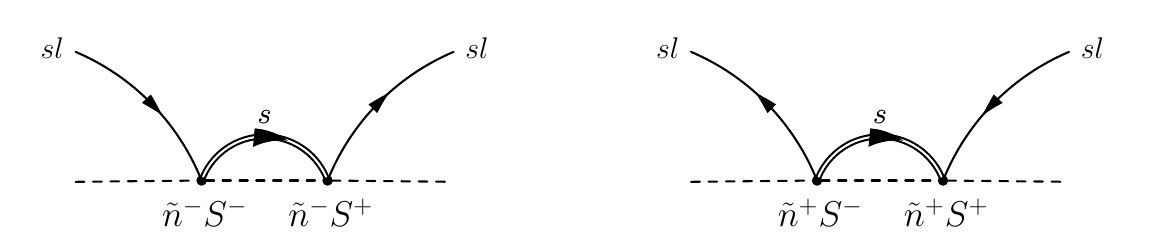}
    \caption{Examples of poor man's scaling ($T$-matrix) diagrams which produce the re\-scaling of the consistent-scheme parallel interaction involving $sl$-sector fermions in the (multi) two-channel Kondo Hamiltonian.}
    \label{fig:consistentJz}
\end{figure}

This can best be seen within the poor man's scaling formalism \cite{Ljepoja2024b}. The diagrams that are involved in the creation of this ``new'' parallel $sl$-sector scattering are shown in Fig.~\ref{fig:consistentJz}. They show how, even if we did not start with the consistent parallel interaction but used the conventional one instead (\textit{i.e.}, having only $s$-sector fermions scatter on the vertex), the re\-scaling of the $J_{z}$ vertex involves processes with $sl$-sector fermions in the external legs. Indeed, starting from the consistent spin-flip and conventional parallel interaction, a new Hamiltonian term is created at the second order that now needs to be included in the original Hamiltonian to see how it feeds back into the flow and make the whole procedure consistent. In such a way one arrives at the consistently refermionized parallel interaction without doing it explicitly (and having to worry about refermionizing $\tilde{n}$'s \cite{Ljepoja2024a}). In the CS calculation, the need for the introduction of an additional interaction term manifests as an inconsistency in the set of equations for the beta-function coefficients. Namely, the RG-consistency equations given by Eq.~(\ref{eq:consistency}) will no longer be satisfied if one uses this \textit{semiconsistent} scheme, where the spin-flip interaction is \textit{consistent} but the parallel one is \textit{conventional}. The latter we rewrite as
\begin{equation}
    H_{K}^{z} = J^{}_{z} \big ( \tilde{n}^{+}_{l,\alpha} + \tilde{n}^{-}_{l,\alpha}\big ) S_{z} \bar{\psi}_{s,\alpha}\psi_{s,\alpha}
\end{equation}
where we have used the property of the $n$-twiddles that $\tilde{n}^{+}_{l,\alpha} + \tilde{n}^{-}_{l,\alpha} = 1$ to be able to make contractions with the consistent spin-flip part of the Hamiltonian.    
The inconsistencies appearing in the system of CS equations are a consequence of the fact that $P_{(4,0)}^{\perp}$ will have a consistent value, since the spin-flip vertices are refermionized consistently, while $P_{(3,0)}^{\perp z}$ will have a conventional value, since $J_{z}$ was refermionized in a conventional way. One needs to introduce an additional ($sl$-sector) interaction to fix the RG consistency. Taking inspiration from the  PMS formalism, it becomes obvious that this additional interaction is the one that scatters $sl$-sector fermions on the $J_{z}$ vertex. More precisely, in order to have the RG-consistency equations satisfied, the parallel part of the Kondo interaction, in a refermionized language, would need to be of the form
\begin{equation}
\label{eq:semicons}
\begin{split}
    H_{K}^{z} = & J^{s}_{z} \big ( \tilde{n}^{+}_{l,\alpha} + \tilde{n}^{-}_{l,\alpha}\big ) S_{z} \bar{\psi}_{s,\alpha}\psi_{s,\alpha} \\
    &+ J^{sl}_{z} \big ( \tilde{n}^{+}_{l,\alpha} - \tilde{n}^{-}_{l,\alpha}\big ) S_{z} \bar{\psi}_{sl,\alpha}\psi_{sl,\alpha}
    \end{split}
\end{equation}
where we have allowed for $s$- and $sl$-sector fermions to scatter with different coupling constants. With this enlarged coupling-constant space, we can capture simultaneously the conventional ($J_{z}^{s} = J_{z}$, $J_{z}^{sl} = 0$) as well as the consistent ($J_{z}^{s} = J_{z}^{sl} = J_{z}$) versions of the $H_{K}^{z}$ interaction. Moreover, the introduction of the necessary additional interaction has turned the semi-consistent model into an RG-consistent one. It is now natural to ask if $J_{z}^{s} = J_{z}^{sl}$ is where the model will flow to under the RG process 
\footnote{Let us remark that the earlier work on the Toulouse limit of the two-lead Kondo model \cite{shah2016,*bolech2016} was done along the same \textit{mixed-BdB} considerations as for the semi\-consistent model discussed here. The fact that ``semi\-consistent'' flows toward ``consistent'', validates that approach. (Notice also that these issues are not present in the case of the Toulouse limit of the two-channel Anderson impurity model \cite{bolech2006a,*iucci2008} due to the differences in how the impurity couples to the bulk.)}. 
The corresponding CS equation is of the form
\begin{equation}
\bigg ( -\frac{\partial}{\partial \mathrm{ln}(\tilde{\omega})}+\beta^{s}_{z} \frac{\partial}{\partial \mathrm{g}^{s}_{z}}+\beta^{sl}_{z} \frac{\partial}{\partial \mathrm{g}^{sl}_{z}}+\beta_{\perp}\frac{\partial}{\partial \mathrm{g}_{\perp}} \bigg ) \ \Sigma^{i} \ = \ 0
\label{eq:CS2}
\end{equation}
All the RG-consistency equations are now automatically satisfied and the beta functions can be calculated in an analogous fashion as in the previous sections. However, the system of equations turns out to be under\-determined with the introduction of the $J^{sl}_{z}$ interaction. One way to increase the number of equations is by using the \textit{consistent and spin-isotropic limit}. This limit is defined as $g^{s}_{z} = g^{sl}_{z} = g = g_{\perp}$, and it provides six additional equations. As it turns out, these additional equations are still not enough to find the three beta functions. Additional information about the beta function is inferred from the PMS procedure. Namely, it can be seen in PMS that there will not be any $J^{3}_{z}$ terms in the beta function. This can be seen by noticing that the wave-function renormalization when $J^{s}_{z} \neq J_{z}^{sl}$ is given by
\begin{equation}
W=1- \big(J^{s}_{z} \big)^2 \rho^2_{0} \frac{|\delta D|}{2 D}  - \big(J^{sl}_{z} \big)^2 \rho^2_{0} \frac{|\delta D|}{2 D}  
\end{equation}
(where we have, for the sake of brevity, put $J_{\perp} \rightarrow 0$; which is enough in this case, since we are interested in only the behavior of the parallel interaction). New coupling constants obtained after reducing the cutoff and following the PMS procedure up to the third order are given by
\begin{equation}
\begin{split}
& J^{s \ \prime}_{z} \simeq J^{s}_{z}+  \big (J^{s}_{z} \big )^3 \rho^2_{0} \frac{|\delta D|}{2D} +  J^{s}_{z} \big (J^{sl}_{z} \big )^2 \rho^2_{0} \frac{|\delta D|}{2D} \\
& J^{sl \ \prime}_{z} \simeq J^{sl}_{z}+  \big (J^{sl}_{z} \big )^3 \rho^2_{0} \frac{|\delta D|}{2D} +  J^{sl}_{z} \big (J^{s}_{z} \big )^2 \rho^2_{0} \frac{|\delta D|}{2D}
\end{split}
\end{equation}

\begin{figure}[b]
    \centering
    \includegraphics[width=0.4\textwidth]{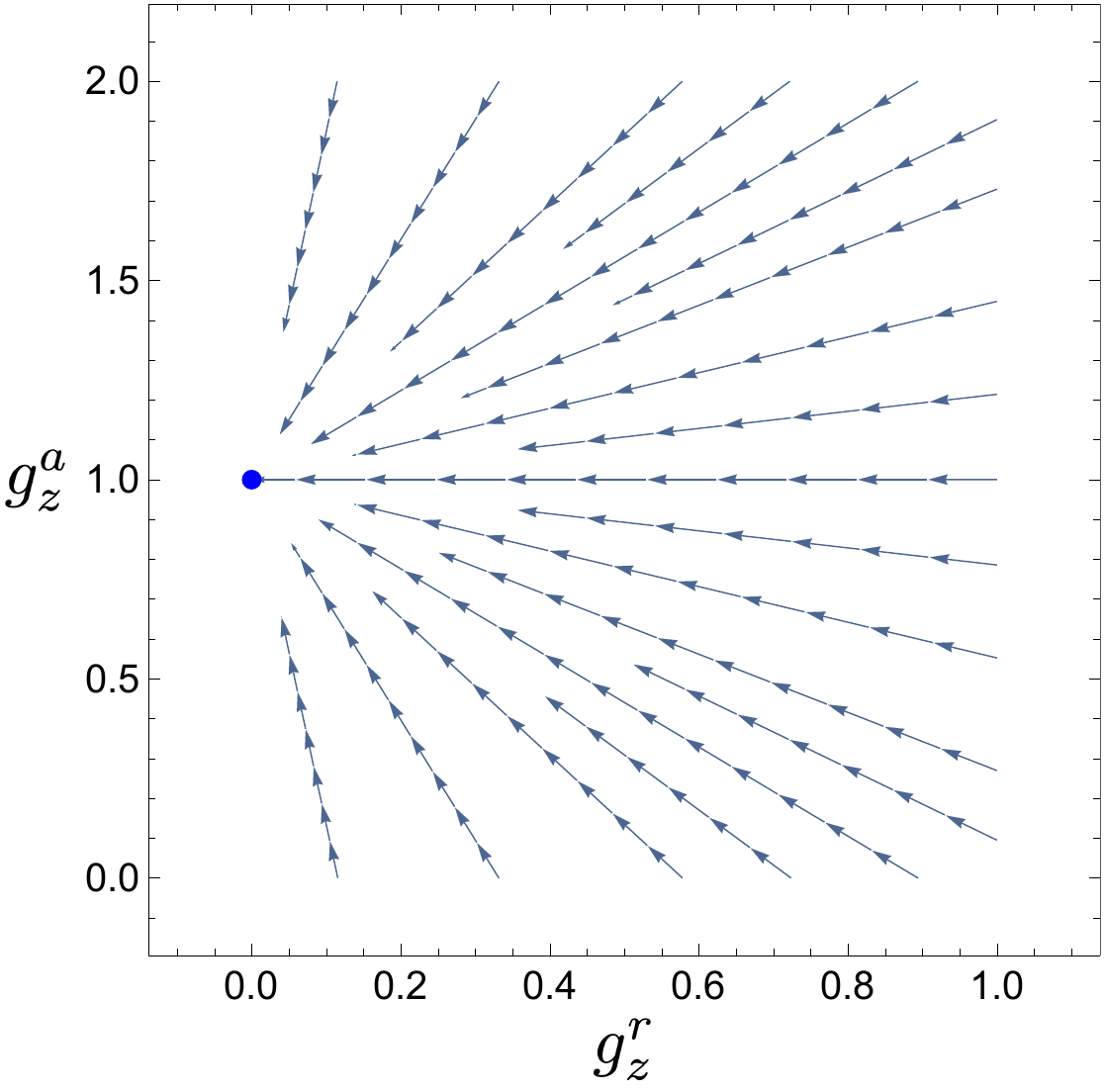}
    \caption{Third-order flow diagram of the coupling constants $g^{a}_{z}$ and $g^{r}_{z}$ for the fixed-point value of the spin-flip coupling constant ($g_{\perp}\!=\!1/M$), plotted for $M\!=\!K/2\!=\!1$. One can see that the RG flow is toward the consistent-model fixed point ($g^{r}_{z}\!=\!0$).}
    \label{fig:semiconsistent}
\end{figure}

In order to obtain the correct flow, one needs to multiply the new coupling constants with the wave-function renormalization. So the flow, up to third order, is actually given by
\begin{equation}
\begin{split}
& W J^{s \ \prime}_{z} \simeq J^{s}_{z}\\
& W J^{sl \ \prime}_{z} \simeq J^{sl}_{z}
\end{split}
\end{equation}
and, therefore, it does not have any $J_{z}^3$ terms and we are justified in demanding that the beta functions obey this also for the CS-based calculation. This additional constraint is enough to solve the algebraic system and obtain the full RG flow
\begin{equation}
\begin{split}
\beta_{\perp} &=  -\frac{1}{2}\ g_{\perp}  \left[ g^{s}_{z} +  g^{sl}_{z} \right] + \frac{M}{2} g_{\perp}^3 + \frac{M}{4} g_{\perp} \left[ \big ( g^{s}_{z} \big)^2 + \big ( g^{sl}_{z} \big)^2 \right] \\
\beta^{s}_{z} &= -\ g_{\perp}^2+ M g_{\perp}^2 g^{s}_{z}\\
\beta^{sl}_{z} &= -\ g_{\perp}^2+ M g_{\perp}^2 g^{sl}_{z}
\end{split}
\end{equation}

One can see that the coupling constants $g^{s}_{z}$ and $g^{sl}_{z}$ appear symmetrically in the flow of $g_{\perp}$ and flow in the same way to the same fixed-point value. This means that, even though we might start with two different values for them, the RG flow will ultimately lead them to a re\-normalized model with  $J^{s}_{z}\!=\!J^{sl}_{z}$, (\textit{i.e.}, the consistent model with $J_{z}\!=\!1/M$). To examine the flow further, it may be more instructive to make a change of variables and introduce $g_{z}^{a}\!=\!\frac{1}{2} (g_{z}^{s} + g_{z}^{sl})$ and $g_{z}^{r}\!=\!g_{z}^{s} - g_{z}^{sl}$. In this way the consistent model is recovered for $g_{z}^{r} \rightarrow 0$. The beta functions for the redefined coupling constants are
\begin{equation}
\begin{split}
\beta_{\perp} &=  -\ g_{\perp}  g^{a}_{z} + \frac{M}{2} \left[ g_{\perp}^3+ g_{\perp} \big ( g^{a}_{z} \big)^2 \right] +\frac{M}{8} g_{\perp} \big ( g^{r}_{z} \big)^2 \\
\beta^{a}_{z}&= -\ g_{\perp}^2+ M g_{\perp}^2 g^{a}_{z}\\
\beta^{r}_{z}&= M g_{\perp}^2 g^{r}_{z}
\end{split}
\end{equation}

By analyzing these flow equations, one can immediately see that $g^{r}_{z}$ is \textit{marginally irrelevant} and will flow towards zero (\textit{i.e.}, to the consistently compactified model). Therefore, when introducing the additional interaction one can as well take $J^{s}_{z} = J^{sl}_{z} = J_{z}$ and thereby obtain the consistently refermionized model that is RG stable (since $g^{r}_{z}\!=\!0$ is a stable fixed hyperplane).

In Fig.~\ref{fig:semiconsistent} we show the flow of $g^{a}_{z}$ and $g^{r}_{z}$ for the fixed-point value of $g_{\perp} = 1/M$ and $M=1$. One can see from the figure that, no matter what the level of asymmetry between $J^{s}_{z}$ and $J_{z}^{sl}$ is, the model flows toward the consistent-scheme limit. This is also true for any other initial value of $g_{\perp}$ one chooses; $g^{r}_{z}$ always flows to zero. These considerations thus provide an alternative way of motivating the consistently refermionized term of the parallel (no-spin-flip) part of the Kondo interaction.

\section{Conclusion and Summary}

In the first article of this triptych \cite{Ljepoja2024a}, we set out to revisit the compactification of Kondo-type quantum impurity models. When using a methodical BdB-based approach to compactification, one could expect that the procedure results in an exact mapping. We showed that in order to achieve such a one-to-one correspondence between the original and the compactified models it is essential that we use the consistent BdB formalism as against the conventional one \cite{shah2016,*bolech2016}, and in doing so we also developed the consistent formalism further. We formulated the way to treat the 
$z$-axis portion of the Kondo coupling term consistently, which involved additional subtleties as compared to the spin-flip part that we had studied in our past work. The step of debosonization required making guided choices in handling the $n$-twiddle factors while reconstructing fermion operators. These choices are corroborated by the RG analysis of a more general class of models that one can obtain within the \textit{consistent}-BdB treatment of the spin-flip terms by judiciously combining poor man's scaling and field-theoretic treatments, as we discussed in the preceding section. A second key extension in the consistent BdB formalism was the discovery of an additional consideration in the consistent treatment of the $n$-twiddle operators necessitated in dealing with diagrams containing fermionic loops: the importance of keeping track of fermion histories and requiring that $n$-twiddles satisfy idem\-potence and co-nilpotence only within each independent history \cite{Ljepoja2024a,Ljepoja2024b}.

In contrast with the earlier studies of boundary field theories that compared the different BdB schemes \cite{shah2016,*bolech2016}, the most unexpected result of the present work is perhaps the seemingly \textit{unreasonable} effectiveness of the \textit{conventionally} compactified (multi) two-channel Kondo model. We trace it to the restrictions in the local electron-impurity scattering process due to the spin algebra of the impurity. As it happens, that turns out to suffice to make the \textit{conventional} scheme recover fully the first two orders of the beta functions (usually considered to be the only \textit{universal} parts); this was one of the highlights of the second article in this series that presented in detail the corresponding poor man's scaling calculations \cite{Ljepoja2024b}. We further saw here that similar restrictions to the electron-impurity scattering are present (to all orders) in the subset of self-energy diagrams that define the so-called (mainstem) large-$M$ limit of the original model. As a result the beta functions obtained using the conventionally compactified model are exact in that limit (even to all orders, going beyond the universal aspects of the physics).

Here we also presented an extended RG calculation that builds upon the basic framework but takes it conceptually further with ideas inspired in effective field theories \cite{Schwartz,Manohar2020,*[{see also in the same volume~}][{}]Neubert2020} and (hyper)-asymptotic analysis \cite{Berry1990,*Berry1991}. It demonstrates how using a cutoff-dependent beta function one can capture (and re\-sum) not only the logarithmic divergences and cutoff-independent terms, but also the like contributions (constant and logarithmic) within the $1/D$ corrections to the self-energy (\textit{i.e.}, within its \textit{subdominant} terms). The leading orders of these corrections can be argued to be universal in a similar (field-theoretic) sense as the leading terms in the standard beta function are, and they capture the universal aspects of the Kondo crossover (farther away from the infrared fixed point). This corresponds to a \textit{minimal} inclusion of microscopic details via a single physical scale identified with the cutoff of the Kondo-type model (given, for instance, by the Schottky scale of a parent Anderson-type impurity model \cite{bolech2002,*bolech2005a}) that indicates when the low-energy description stops sufficing. Those corrections are thus reflected in a finite-temperature regime of the free energy and thermodynamic quantities close to but above the Kondo temperature. They are usually of limited interest due to the difficulty in isolating them experimentally (or even calculate them theoretically) but are interesting in the context of the conventional vs.~consistent comparisons. After the inclusion of the $1/D$ corrections, both compactifications start to differ from each other at higher temperatures even for the extended-sense-universal equilibrium aspects of the physics (that are captured by minimal models with finite cutoffs but still do not depend on the cutoff value, and are also independent of further microscopic details of particular physical realizations).

In conclusion, our proposal for the \textit{consistent} compactification of the (multi) two-channel Kondo model is argued to be exact. It passes all the comparison tests we made across this series of three articles: different exact results in solvable limits were in matching accordance with the original model \cite{Ljepoja2024a}, the universal part of the RG flow coincided \cite{Ljepoja2024b}, and even the nonuniversal aspects of the flow agreed as we saw in here. On the other hand, although the \textit{conventional} compactification is not exact, we found that it constitutes an excellent approximation that correctly captures many aspects of the physics. 

At this point, let us pause and zoom out to place this work in the bigger context of using bosonization to study low-dimensional correlated systems. Coming from the motivation of generalizing bosonization to nonequilibrium problems \cite{shah2016,*bolech2016}, it was discovered that for the case of a tunneling junction there was a discrepancy between the direct calculations and those obtained after a BdB procedure. Its diagnosis revealed that the conventionally implemented BdB procedure violated the symmetries of the original problem. A critical and careful deconstruction of the subtle technicalities of the BdB procedure led to the emergence of the consistent BdB formalism \cite{shah2016,*bolech2016}. Note that, since a standard unfolding procedure was used to map the zero-dimensional problem to a model of one-dimensional chiral fermions with linear dispersion, BdB was expected to be exact for the whole many-body spectrum. As a result, the junction study offered an ideal and controlled situation for the purpose of systematic comparisons with direct results. This is also the case for the Kondo-type models comprehensively studied in the subsequent papers, including this one \cite{shah2016,*bolech2016,Ljepoja2024a,Ljepoja2024b}, and for many other zero-dimensional and boundary models. Hence we can be sure that the source of discrepancy lies in the BdB steps and not in any of the presteps, such as linearization, which might be necessary in some other cases (see footnote [21] of Ref.~\onlinecite{Ljepoja2024a}). That also implies that, once the linearization step is performed, the implications of our finding of symmetry violations will need to be evaluated on a case-by-case basis. 

It follows now by extension that the matching of conventional-BdB with consistent-BdB and direct results for the next-to-lowest-order RG beta-function terms, despite the nonconservation of originally conserved quantities, is serendipitous. 
Furthermore, the discrepancies start to show at the lowest nontrivial order in perturbation and can be manifest, depending on the problem, not only in the finite-temperature and nonequilibrium physics, but also in the universal regime of the low-energy thermodynamics.
It is thus imperative to be cognizant of the consistent BdB procedure to reliably avail the advantages of one of the ubiquitous techniques to study low-dimensional correlated systems that can also be used far from equilibrium. 

Since the conventional scheme can be obtained as a mean-field type approximation of the consistent one, we can envision the possibility of refining the approximation by including fluctuation corrections. This could be an interesting avenue for further developments that would help clarify open questions like the mechanisms for breaking of universality in out-of-equilibrium settings (like steady-state regimes in quantum dots). Further on, it would be interesting to revisit the case of multiple impurities and, away from the dilute limit, investigate the possible implications for the study of low-dimensional disorder systems \cite{Giamarchi} (for both magnetic and nonmagnetic impurities). Finally, pursuing additional studies of the formal mathematical structure of consistent BdB might open up another set of interesting avenues for inquiry. One example of such directions would be to systematically compare consistent BdB with the non-Abelian bosonization framework.

\acknowledgements  
We acknowledge discussions about field-theoretic renormalization group formulations and beta functions with several colleagues, in particular N.~Andrei and L.\,C.\,R.~Wijewardhana. We also acknowledge the hospitality of the International Center for Advanced Studies (ICAS; UNSAM, Argentina) where part of this work was done.

\appendix

\section{Popov-Fedotov diagrammatic technique}

\label{Apdx:PF}
In our analysis of the Kondo-model self-energy, it is convenient to represent the impurity spin in the diagrammatic expansion using some auxiliary-particle technique that enables one to apply Wick's theorem \cite{Wick1950}. We opted for using Popov-Fedotov (PF) pseudo-fermions \cite{popov1988,Veits1994,*Kiselev2001}, a constraint-free option where each impurity spin state is represented by a fermionic degree of freedom. 

The basic idea of this technique is as follows. To start, one writes the impurity spin as
\begin{equation}
    S_\mathrm{imp}=\sum_{\mu, \nu} \eta^{\dagger}_{\mu} \sigma_{\mu \nu} \eta_{\nu}
\end{equation}
where the $\eta$'s are auxiliary fermionic degrees of freedom that satisfy standard anticommutation relations,
\begin{equation}
   \{ \eta^{\dagger}_{\mu} , \eta_{\nu} \} = \delta_{\mu\nu}
\end{equation}
These fermionic degrees of freedom describe the dynamics of the spin impurity and enable us to use standard perturbative field theoretical methods for finding the flow of the coupling constants. On the other hand, the downside of this type of fermionic spin representations is the enlargement of the impurity Hilbert space with the appearance of unphysical states that would also contribute to the partition function. One needs to remove those spurious contributions, and different auxiliary-particle techniques tackle that in different ways (often, introducing constraints). The PF approach is by including a purely imaginary chemical potential.

For concreteness, we shall focus here on the spin-$\frac{1}{2}$ case. (The spin-1 case can be handled very similarly; larger spins and other algebras have also been implemented in the literature, but required nontrivial generalizations.) There are in total four states per impurity in the PF basis: a spin-singlet state with no fermions, two states with a single fermion (a spin-doublet state), and another spin-singlet state with two-fermion occupancy (with opposite spin orientations). The partition function for some PF-enlarged model is thus given by $Z=Z_{0}+Z_{1}+Z_{2}$, where we define
\begin{equation}
\begin{split}
   Z_{0} & = \mathrm{Tr} [\delta(N)\,e^{-\beta (H-\mu N)}] 
           = \mathrm{Tr} [e^{-\beta H}] \\
   Z_{1} & = \mathrm{Tr}[\delta(N-1)\,e^{-\beta(H-\mu N)}]  
           = e^{\beta\mu} \ \mathrm{Tr}[e^{-\beta H}]\\
   Z_{2} & = \mathrm{Tr}[\delta(N-2)\,e^{-\beta (H-\mu N)}] 
           = e^{2\beta\mu} \ \mathrm{Tr}[e^{-\beta H}]\\
\end{split}
\end{equation}
with $N$ being the number of PF fermions at the impurity site and $\mu$ an as-yet-unspecified chemical potential for those fermions, while $H$ can be our Kondo Hamiltonian in which the spin operators have been replaced by the auxiliary fermionic degrees of freedom. From the original model, the only physical contribution to the partition function is $Z_{1}$; the other two contributions ($Z_{0}$ and $Z_{2}$) are unphysical and come from the empty and doubly occupied PF-fermion states. The PF \textit{trick} consists in eliminating these unphysical contributions to the partition function by introducing an imaginary chemical potential, say $\mu = \frac{i}{2}\frac{\pi}{\beta}$ (or the opposite sign choice), for the PF fermions. In that way, the unphysical contributions cancel each other (we have $Z_{0}+Z_{2} = 0$), and the only remaining contribution to the partition function is the physical one ($\mathrm{Tr}[e^{-\beta H}] = e^{-\beta\mu} Z_{1} = -iZ$). 

We shall also refer to the PF fermions with this imaginary chemical potential as the pseudo-fermions. These new pseudo-fermions are treated the same way as any other fermionic degree of freedom in the theory. The free Matsubara propagator is simply given by
\begin{equation}
G^\textsc{pf}_{0} (\omega_m)= \frac{1}{\mu - i \omega_m}
\end{equation}
where the imaginary chemical potential can be interpreted as shifting by $\frac{1}{2}$ the index of the Matsubara frequencies from their standard fermionic values, $\omega_m|_\mathrm{Fermi}=\frac{\pi}{\beta}m|_{m\in\mathbb{Z}_\mathrm{odd}}$, and thus altering all frequency summations. Due partly to this, and for high orders in perturbation theory, we will find it more convenient to adopt a Fourier-transformed representation, carry out imaginary-time integrals instead of frequency summations, and then Fourier-transform back the final results for analytic continuation to real frequencies.

\section{The $sl$-fermion self-energy}
\label{Apdx:slSelfEnergy}

When doing the CS calculation of the RG flow in the indirect schemes, we have a choice of whether we use the $s$- or $sl$-fermion self-energy. As it turns out, in the conventional scheme that choice needs to be made carefully, since the $sl$-fermion self-energy does not by itself provide enough information to enable the determination of the beta functions. The reason is that, in that scheme, the $J_{z}$ vertex scatters only $s$-sector fermions, limiting the number of possible diagrams one can make. As a result, there is no $J_{z}^2$ contribution to the self-energy. That is evident, since there is no way one can make a self-energy diagram with external (amputated) legs being $sl$-sector fermions by using two $J_{z}$ vertices that scatter only $s$-sector fermions. In addition, the multiplicative factors are going to be different (as compared to those in the $s$-fermion self-energy) for any  diagrams that have one or more $J_{z}$ vertices in them. Taking all of this into account, one arrives at the same result as in Eq.~(\ref{eq:SEc}), only in this case the $P$ coefficients will have different values, namely,
\begin{align}
&P^{\perp}_{(2,0)} = 4  &  &P^{z}_{(2,0)} = 0 \nonumber\\
&&  &P^{\perp z}_{(3,0)} = -16\nonumber\\
&P^{\perp}_{(4,0)} = 32  &  &P^{\perp z}_{(4,0)} = 32 \nonumber \\
&P^{\perp}_{(4,1a)} = 16 \big [ \ln(2)-1 \big ] & \quad &P^{\perp z}_{(4,1a)} = 16 \big [ \ln(2)-1 \big ] \nonumber\\
&P^{\perp}_{(4,1b)} = 16   &  &P^{\perp z}_{(4,1b)} = 16 \nonumber\\
&P^{\perp}_{(4,0a)} = -\pi^3   &  &P^{\perp z}_{(4,0a)} = \pi^3 
\end{align}

Two noticeable differences are thus the lack of $P^{z}_{(2,0)}$, and $P^{\perp z}_{(4,0)}$ being negative. Both of those differences are direct results of not having $sl$-fermion scattering on the $J_{z}$ vertex. This, on the other hand, is a direct consequence of the usage of the conventional BdB scheme for refermionization. The beta functions ans\"atze are still the same, Eq.~(\ref{eq:beta0}), as they were when using the $s$-fermion self-energy. But the resulting equations to calculate the coefficients in the ans\"atze are, instead,
\begin{align}
        a_{1} &= \frac{P_{(3,0)}^{\perp z}}{2 P_{(2,0)}^{\perp}}  & \quad \nonumber\\
 a_{2} &= \frac{M P_{(4,1b)}^{\perp}}{2 P_{(2,0)}^{\perp}}  \ & \quad b_{1,2,3,4} &=\,???? \\
 a_{3} &= \frac{M P_{(4,1b)}^{\perp z}}{2 P_{(2,0)}^{\perp}} \ & \quad \nonumber
\end{align}
where we have used question marks to indicate the coefficients that remain undetermined. We can also see, by comparing to the set of equations found before, in Eq.~(\ref{eq:coeffD}), that the lack of the second-order $J_{z}^2$ contributions breaks the connection that existed between the $a_{1}$ and $b_{1}$ coefficients (as well as $a_{3}$ and $b_{3}$).
In fact, the entire flow of the $J_{z}$ coupling constant stays undetermined. 

In addition, we again have the consistency equations
\begin{equation}
    \begin{split}
     2 P_{(3,0)}^{\perp z} a_{1} & = 2 P_{(4,0)}^{\perp z} - P_{(3,0)}^{\perp z} b_{2}\\
     P_{(3,0)}^{\perp z} b_{1} & = 2 P_{(4,0)}^{\perp} \\
    \end{split}
    \label{eq:consis}
\end{equation}
One can use those equations to find some of the unknown coefficients. Therewith, one could find $b_{1}$ from the second consistency equation. But that would mean that in order to calculate the beta function to second order in the coupling constant one would have to go to a higher order in the self-energy than is needed when using the $s$-fermion self-energy. Additionally, and more importantly, we arrive at the value $b_{1}=-4$, which disagrees not only with the direct scheme, but also with the conventional one using the $s$-fermion self-energy. On the other hand, since we know $a_{1}$, the coefficient $b_{2}$ can be calculated from the first of the consistency equations. One finds $b_{2}=0$, which is as expected. However, the isotropic limit is no longer satisfied for such values of $b_{1}$ and $b_{2}$. The rest of the coefficients of $\beta_{z}$ remain undetermined. From all of this, it is evident that if one wants to determine the ``correct'' RG flow, in the conventional language one needs to choose the $s$-fermion self-energy for the CS-based calculation. Otherwise one runs into an undetermined and inconsistent system of equations. On the other hand, in the consistent scheme, where $s$ and $sl$ sectors enter on an equal footing, there is no need to work with only the $s$ self-energy. In fact, both $s$- and $sl$-fermion self-energies are equal in the consistent scheme, producing the same coefficients and the same equations for the beta functions.

\section{Beta function of the spin-anisotropic Kondo model beyond third order}
\label{Apdx:anisotropic4th}

Moving away from the spin-isotropic limit, the system of linear constraints arising from the CS equation becomes under\-determined. This continues to be the case even when one includes (as we did for the lower-order calculation) additional equations coming from the spin-isotropic limit. One therefore needs even more equations added to the system in order to have a single solution.

The procedure for finding the beta functions is the same as in the isotropic case. To get the fourth-order RG flow, we assume beta-function ans\"atze of the form
\begin{alignat}{2}
\label{eq:betaAns2}
\beta_{\perp}(\mathrm{g}_{z},\mathrm{g}_{\perp}) & = a_{1} \mathrm{g}_{z}\mathrm{g}_{\perp}+ a_{2}  \mathrm{g}_{\perp}^3+ a_{2M} M \mathrm{g}_{\perp}^3 +  a_{3}\mathrm{g}_{\perp} \mathrm{g}_{z}^2 \nonumber \\
&\quad + \ a_{3M} M \mathrm{g}_{\perp} \mathrm{g}_{z}^2 +  a_{4}\mathrm{g}_{\perp} \mathrm{g}_{z}^3+ a_{5} \mathrm{g}_{\perp}^{3} \mathrm{g}_{z} \nonumber\\
& \quad  \\
\beta_{z}(\mathrm{g}_{z},\mathrm{g}_{\perp}) & = b_{1} \mathrm{g}_{\perp}^2 + b_{2}\mathrm{g}_{z}^2+  b_{3}\mathrm{g}_{\perp}^2 \mathrm{g}_{z} +  b_{3M} M \mathrm{g}_{\perp}^2 \mathrm{g}_{z} \nonumber\\
&\quad + \ b_{4} \mathrm{g}_{z}^3+ b_{4M} M \mathrm{g}_{z}^3 +b_{5} \mathrm{g}_{\perp}^2\mathrm{g}_{z}^2 + b_{6} \mathrm{g}_{\perp}^4 \nonumber
\end{alignat}
where we separated explicitly the channel-dependent from the channel-independent parts of the beta functions. We do this for the same reason as in the spin-isotropic case: we computed only the channel-independent fifth-order contributions  (the non-fermion-loop diagrams) to the self-energy. Therefore, we have to do an expansion of the CS equation in both coupling constant and channel number, and make sure that we keep only equations that are of order lower than $\mathrm{g}^5 M$. So, we collect coefficient pre\-factors in front of each $\mathrm{g}^{a} M^{b}$ power (where $a+b\le5$) and get a set of linear constraints on the ansatz coefficients. The imaginary part of the retarded self-energy, to which we are applying the CS equation, is going to be the same as in Eq.~(\ref{eq:SEd}) with the terms in Eq.~(\ref{eq:5thDcont}) added to it. This produces a system of equations, but we still have the problem of it being under\-determined. We can fix this by introducing additional conditions. One such a condition is what we call the ``perturbative Toulouse limit''. In it, we assume that $\mathrm{g}_{z}$ does not flow along the $\mathrm{g}_{z}\!=\!\mathrm{g}^{\star}$ line, where $\mathrm{g}^{\star}$ is a fixed point of the spin-isotropic model (that is not under\-determined). In other words, one asks that
\begin{equation}
\beta_{z}(\mathrm{g}_{z} \rightarrow \mathrm{g}^{\star}, \ \mathrm{g}_{\perp}) = 0
\end{equation}

With the help of this additional constraint on the RG flow, one can calculate all of the unknown coefficients in Eq.~(\ref{eq:betaAns2}) and arrive at
\begin{equation}
\begin{split}
\beta_{\perp} \ =  & -g_{z}g_{\perp}+ \frac{M}{2} \ \big ( g_{\perp}^3 +  g_{\perp} g_{z}^2 \big ) + \frac{\pi^2}{4} \ g_{\perp}^{3} g_{z} \\
\\
\beta_{z} \ = & -g_{\perp}^2 + M \ g_{\perp}^2 g_{z} +\frac{\pi^2}{4} \ g_{\perp}^2g_{z}^2 \\
\end{split}
\label{eq:beta_ani_D}
\end{equation}
where we have already made the substitution of $\mathrm{g}_{\perp} \rightarrow g_{\perp}/2$ and $\mathrm{g}_{z} \rightarrow g_{z}/2 $. These are the spin-anisotropic RG-flow equations calculated in the direct scheme. We can repeat the whole calculation in the indirect schemes. The consistent scheme, as expected, gives the same result as the direct one. On the other hand, as we have seen for the case with spin isotropy, conventional compactification produces a different RG flow
\begin{equation}
\begin{split}
\beta_{\perp}  =  & -g_{z}g_{\perp}+ \frac{M}{2} \ \big ( g_{\perp}^3 +  g_{\perp} g_{z}^2 \big )\\
&\qquad \qquad \qquad + \frac{41 \pi^2}{96} \ g_{\perp}^{3} g_{z} - \frac{49 \pi^2}{192} \ g_{\perp} g_{z}^3 \\
\\
\beta_{z}  = & -g_{\perp}^2 + M \ g_{\perp}^2 g_{z} +\frac{11 \pi^2}{64} \ g_{\perp}^2g_{z}^2 \\
\end{split}
\label{eq:beta_ani_D}
\end{equation}
As we can see, the beta functions of all three schemes agree completely up to the third-order expansion (even though there is an unphysical diagram already present), but at fourth order the conventional scheme starts to diverge from the other two. Not only are there different beta function coefficients between them, but there is a whole new term in the conventional-scheme RG flow of $g_{\perp}$ that does not exist in the direct (or consistent) scheme. 

\section{Full beta function of the Kondo model beyond third order}
\label{Apdx:FullBetaIsotropic4th}

In our calculations of the fourth-order RG flow, we have taken into account only the fifth-order \textit{channel-number-independent} contributions to the self-energy. This was done for brevity, because for those contributions we have unphysical diagrams contributing in the conventional scheme, but not for the \textit{channel-number-dependent} ones. The latter contributions to the fifth-order self-energy have been been calculated before in the spin-isotropic limit \cite{gan1994}. We can use those results, in combination with the result we derived, to arrive at the full fourth-order beta function of the (multi) two-channel Kondo model. The self-energy is again of the form as in Eq.~(\ref{eq:SE5thd}), but now instead of ignoring the $P_{(5,1a)}$ and $P_{(5,1b)}$ coefficients, we use their values given by
\begin{equation}
\begin{split}
    & P_{(5,1a)} = -168\\
    & P_{(5,1b)} = 48 [5-3\ln(2)]
\end{split}
\end{equation}
On the other hand, the beta-function ansatz is slightly changed to include a channel-number-dependent fourth-order term
\begin{equation}
\beta(\mathrm{g}) = a_{1} \mathrm{g}^2 + a_{2} \mathrm{g}^3 + a_{2M} M \mathrm{g}^3 + a_{3} \mathrm{g}^4 +a_{3M} M \mathrm{g}^4 
\end{equation}
The system of equations one arrives at (via the CS procedure) with this higher-order self-energy and beta-function ansatz is the same as in Eqs.~(\ref{eq:eqD5th}) and (\ref{eq:eqD5thc}), with the addition of an extra linear constraint on the coefficients and an additional RG-consistency requirement. The former addition, is the relation needed for determining the $a_{3M}$ coefficient of the beta function
\begin{equation}
a_{3M} = \frac{P_{(5,1b)} - 4 P_{(4,1a)}a_{1}}{2 P_{(2,0)}} = 8 [1+\ln(2)]
\end{equation}
while the latter additional RG-consistency condition is
\begin{equation}
P_{(5,1b)} = 2 P_{(4,1b)}a_{1} + \frac{3}{2} P_{(3,0)} a_{2M}
\end{equation}
This is the same situation one would find using the conventional scheme for the compactification of the model. It is as expected, since the fifth-order channel-number-dependent contributions to the self-energy are, as we have discussed in the main text, all \textit{translatable}. Therefore, in the channel-number-dependent part of the fourth-order beta function we do not expect to see any disagreement between schemes. 

Explicitly, the full fourth-order RG flow of the spin-isotropic model is
\begin{equation}
\begin{split}
\beta &= -2 \mathrm{g}^2 + 4 M \mathrm{g}^3 +2 \pi^2 \mathrm{g}^4 + 8 \ [1+\ln(2)] M \mathrm{g}^4\\
&\mapsto -g^2+ M g^3 +\frac{\pi^2}{4}g^4 + [1+\ln(2)] M g^4
\end{split}
\end{equation}
and, as we have established by analyzing the diagrammatic contributions, the fourth-order channel-number-dependent part of the flow is going to be the same across schemes, and the difference in the conventional scheme will be in the channel-number-independent terms only.

\section{Subdominant contribution to the running coupling constant and scaling} \label{Apdx:subdominant}

Here we investigate further the effects of keeping the subdominant terms in the RG analysis. For such a task, it is convenient and more instructive to do it in the spin-isotropic limit, for which all three schemes give the same result for the universal part of the beta function (even in the extended sense, which indicates a very broad applicability of the conventional compactification scheme, since it can also be used to re\-sum the subdominant terms in the self-energy). Having obtained the beta functions including the first subdominant order, one can calculate the running coupling constant. This provides information on how physical quantities scale with temperature and their scaling dimensions. Unlike when dealing with only the dominant part of the RG flow, calculating the running coupling constant is now not so straightforward anymore. The reason is that the differential equation one uses to determine it is no longer separable, since the beta function itself is $\tilde{\omega}$ dependent.  

Explicitly, the differential equation we are hoping to solve is
\begin{equation}
\begin{split}
    \frac{d g}{d \mathrm{\ln}(\tilde{\omega})} & = -\big (1+\tilde{\omega}^2 \big ) \ g^2 + M g^3\\
    & \approx \Delta \bigg( \frac{g}{g^{\star}} \bigg)^2 (g-g^{\star}) - \tilde{\omega}^2 g^2
\end{split}
\label{eq:subdominant}
\end{equation}
where we have have approximated the dominant part of the beta function using its slope $\Delta$ and the fixed point value $g^{\star}$. Basically (cf.~Ref.~\onlinecite{Ljepoja2024b}), we replace the original higher-order-in-$g$ corrections to the dominant flow with a lower-order interpolant that has the same location-of and slope-at $g^{\star}$ as the higher-order beta function does. We can introduce the deviation of the coupling constant from the fixed point as $g=g^{\star}+h$, and keeping only terms that are linear in $h$, the differential equation in Eq.~(\ref{eq:subdominant}) is reduced to a linear and inhomogenous equation for $h$
\begin{equation}
    \frac{d h}{d \tilde{\omega}} - \frac{\Delta}{\tilde{\omega}} h = - \tilde{\omega} (g^{\star})^2 
\end{equation}
This differential equation can be solved using standard methods to obtain its general solution,
\begin{equation}
    g(\tilde{\omega})= g^{\star} + a_{0} \ \tilde{\omega}^{\Delta} - \tilde{\omega}^2 \ (g^{\star})^2/(2 - \Delta)
\end{equation}
where $a_{0}$ is an integration constant that we can fix using the condition that $g(\tilde{\omega} \rightarrow t_{K}) = 2 \ g^{\star}/3$, with $t_{K} \equiv T_{K}/D$ as Kondo temperature divided by the cutoff. This gives us an explicit expression for the running of the coupling constant,
\begin{equation}
\begin{split}
    g(\tilde{\omega}) = g^{\star} - \ &  g^{\star} \ \bigg [ \frac{1}{3} - g^{\star} \ t_{K}^2 / (2 - \Delta) \ \bigg ] \bigg ( \frac{\tilde{\omega}}{t_{K}}\bigg )^{\Delta} \\
    & - \tilde{\omega}^2 \ (g^{\star} )^2 / (2 - \Delta)
    \end{split}
\end{equation}

Comparing this result with the dominant running coupling constant expression \cite{Ljepoja2024b} we see that, besides the $\tilde{\omega}$ dependence that scales with the Kondo temperature, we also have an additional $\tilde{\omega}^2$ dependence that is scaled by some other temperature scale that we call $t^{\star}$ and define it as $t^{\star} = \sqrt{2-\Delta }/g^{\star}$. So, the running coupling constant can be expressed as
\begin{equation}
    g(\tilde{\omega}) = g^{\star} - \xi \left( \frac{\tilde{\omega}}{t_{K}}\right)^{\Delta} - \ \left( \frac{\tilde{\omega}}{t^{\star}} \right)^2
\end{equation}

where $\xi = g^{\star} \ \big (1/3 \ - \ g^{\star} \ t_{K}^2 / (2 - \Delta) \big )$. The physics governed by the scale $t^{\star}$ plays in at higher temperatures than the Kondo temperature (\textit{i.e.}, $t^{\star} > t_{K}$). This means that as one moves towards higher temperatures (away from $t_{K}$) subdominant contributions become gradually more important. Having determined the scaling of the running coupling constant, makes it possible to calculate the scaling of any physical quantity that can be expressed in terms of it. In particular, the scaling of the self-energy itself can be considered. From it, using the fact that the scattering rate due to the Kondo interaction and the imaginary part of the retarded self-energy are connected through $\tau = 1/(2 \mathrm{Im}(\Sigma))$, the scaling of the scattering rate itself can be derived. 
This gives access to the temperature dependence of the resistivity, an important quantity which can be calculated using Kubo's formula (which is an integral in $\tilde{\omega}$ of the scattering rate). Doing that calculation, one obtains an impurity contribution to the resistivity that scales as
\begin{equation}
    \rho(T) \approx \rho_{0}\bigg (\frac{T}{T_{K}} \bigg )^{\Delta} + \rho_{1}\bigg (\frac{T}{T^{\star}} \bigg )^{2}
\end{equation}
where $\rho_{0}$ and $\rho_{1}$ are unspecified numerical coefficients. Besides the low-temperature scaling with the Kondo temperature, one can also see (in the high-temperature deviations) the appearance of the new scale, $T^{\star}$, that the resistivity will be sensitive to at those higher temperatures. While its experimental determination becomes much more difficult, the new energy scale is nevertheless also universal in principle.

%

\end{document}